# Title: The Information Ecosystem of Online Groups with Anti- and Pro-vaccine Views on Facebook

**Authors:** Soojong Kim[1,2]*, Kwanho Kim[3,4]

**Affiliations:**

[1]Cyber Policy Center, Stanford University; Stanford, CA 94305, United States.

[2]Center on Philanthropy and Civil Society, Stanford University; Stanford, CA 94305, United States.

[3]Department of Communication, Cornell University; Ithaca, NY 14853, United States.

[4]Annenberg School for Communication, University of Pennsylvania; Philadelphia, PA 19104, United States.

*Corresponding author. Email: sjkim97@stanford.edu

**Abstract:** Opposition and hesitancy to vaccination have been one of the major threats to global health. Social media sites have been suspected as a breeding ground of misleading narratives about vaccines, but little is known about how pervasive anti-vaccine views are on the world's largest social media, Facebook. Here, we study the prevalence of online groups on Facebook with anti- and pro-vaccine views and the information ecosystem that enables the production and dissemination of anti-vaccine narratives, using the largest collection of Facebook data on this issue to date. The results reveal that 74% of posts each month on average were generated by anti-vaccine groups. It is also shown that anti-vaccine groups had increasingly more relied on relatively credible sources while their posts using low credibility sources were less than 2% and recently decreasing. Furthermore, anti-vaccine groups depended more on their exclusive sources and utilized sources representing more conservative or far-right political views than pro-vaccine groups did. The findings suggest that expansive and targeted interventions are urgently needed to curb the circulation of online narratives against vaccination.

**Main Text:** There has been growing concern about public distrust in science (*1–3*). Especially, opposition and hesitancy to vaccination have been one of the major threats to global health, becoming a significant obstacle in curbing the COVID-19 pandemic that has taken over 4 million lives around the globe (*4–6*). Social media services have been known to fuel distrust and exacerbate the "infodemic" by providing community spaces for online groups spreading misleading and unsubstantiated narratives about vaccines (*5*, *7–10*). It has been suspected that the proliferation of anti-vaccine groups and content is especially severe on Facebook, the world's largest social media site that has more than 3 billion users (*8*, *11–13*). However, evidence is still thin about how pervasive anti-vaccine views are on Facebook and how anti-vaccine groups produce their false narratives on the platform.

Here, we present an analysis of 2,328 online groups that discussed vaccines and vaccinations on Facebook between 2012 and 2020. These groups generated over 1.6 million posts written in English during the period that induced a total of over 700 million shares, comments, and reactions, and their messages were transmitted over 140 billion times since 2018 to Facebook users following them (SM S3). We investigate the prevalence of anti- and pro-vaccine views among them and the information ecosystem that these groups relied on to support their views. This research focused on three critical but largely unanswered questions: How pervasive anti-vaccine views are among online groups discussing vaccines and vaccination on Facebook? How dependent are these groups on information from low credibility sources? How do these groups with different views interact within the broader information ecosystem?

This study leverages the largest observation of online groups with pro- and anti-vaccine views on Facebook, including their recent development during the COVID-19 pandemic. Previous research provided important initial findings. Johnson et al. (*11*) analyzed 441 Facebook pages with anti- and pro-vaccine views, focusing on the connections between them created by the page administrators. Schmidt et al. (*14*) collected data of 243 vaccine-related pages on Facebook and analyzed their volume of content and number of user interactions. Compared with the past research, considerably more online groups were tracked in the current study for a longer time period, which helps us make more reliable platform-level observation and inference (SM S2).

We also expand the approaches of the existing literature by exploring a broader information ecosystem contributing to the production of misleading anti-vaccine narratives. Previous research on online falsehood has focused on how fake news sites or stories were utilized on social media (*15–18*), but we need to have a fuller understanding of online falsehood by examining a wider range of online actors beyond a small set of fake news sources (*3*, *15*, *19*, *20*). Thus, in addition to websites supplying misinformation (*15*, *17*, *18*), the present research explores how vaccine groups interact with other types of information sources known to be crucial in investigating social phenomena through the online space, including news sites that have been viewed as relatively credible sources of information (*15*, *17*, *21–23*), information sources run by government authorities (*24–26*), and other social media platforms, such as Twitter and Instagram (*27*, *28*).

In the current study, a "vaccine group" refers to a community of Facebook users who created, consumed, and engaged with the content of a Facebook page or a Facebook group discussing vaccines and vaccination. A vaccine group can refer to people involved with either a Facebook page or a Facebook group. We identified each group's view on vaccines and vaccination, and anti-vaccine groups were defined as vaccine groups spreading misleading narratives that emphasize the negative consequences of vaccinations and encourage people to reject and avoid vaccines. See S1 for details.

We identified 2,328 vaccine groups on Facebook during the period between January $1^{st}$, 2012, and November $30^{th}$, 2020. Each group was reviewed and classified by multiple coders into either anti-vaccine, pro-vaccine, or mixed/neutral groups. See S2 for details.

Anti-vaccine views had been dominant among online groups discussing vaccines on Facebook. Also, an anti-vaccine group created more content and stayed active for a longer period of time than a pro-vaccine group on average. First, when averaged over the 107 months, 74.1% (*SD* = 6.2%) of all vaccine posts were generated by anti-vaccine groups each month. The proportion of posts created by anti-vaccine groups steadily declined over time as shown in Fig. 1A. The total number of posts produced by vaccine groups had increased since 2012, with noticeable surges that coincide with three major infectious disease outbreaks in the U.S. (See S3.1). Second, the majority of vaccine groups held anti-vaccine views. Anti-vaccine groups accounted for 58.4% (*SD* = 2.9%) of all active vaccine groups each month. The total number of active groups has increased 5.4 times between January 2012 and November 2020 (Fig. 1B). Third, an active anti-vaccine group has created 2.4 times more posts in a year on average than an active pro-vaccine group (Fig. 1D). Lastly, when the lifetime of a group, i.e., the time between its first and last content publication, was computed, the average lifetime of anti-vaccine groups was 3.3 years (*SD* = 2.9 years, *N* = 1,083), and it was 1.5 times greater than that of pro-vaccine groups (*M* = 2.1 years, *SD* = 2.7 years, *N* = 1,206), as displayed in Fig. 1E. See S3 for details.

Considering Facebook users' engagement with the vaccine content, we found that anti-vaccine groups attracted more user engagement than pro-vaccine groups, but there was a noticeable change recently (Fig. 1C). The content generated by anti-vaccine groups had been shared more than pro-vaccine groups' until February 2020, but pro-vaccine groups started attracting more shares than their counterpart since the early stage of the COVID pandemic. Similarly, the proportions of comments and reactions received by anti-vaccine groups also plummeted in 2020 after a long-term decline between 2012 and 2019. See S3 for details.

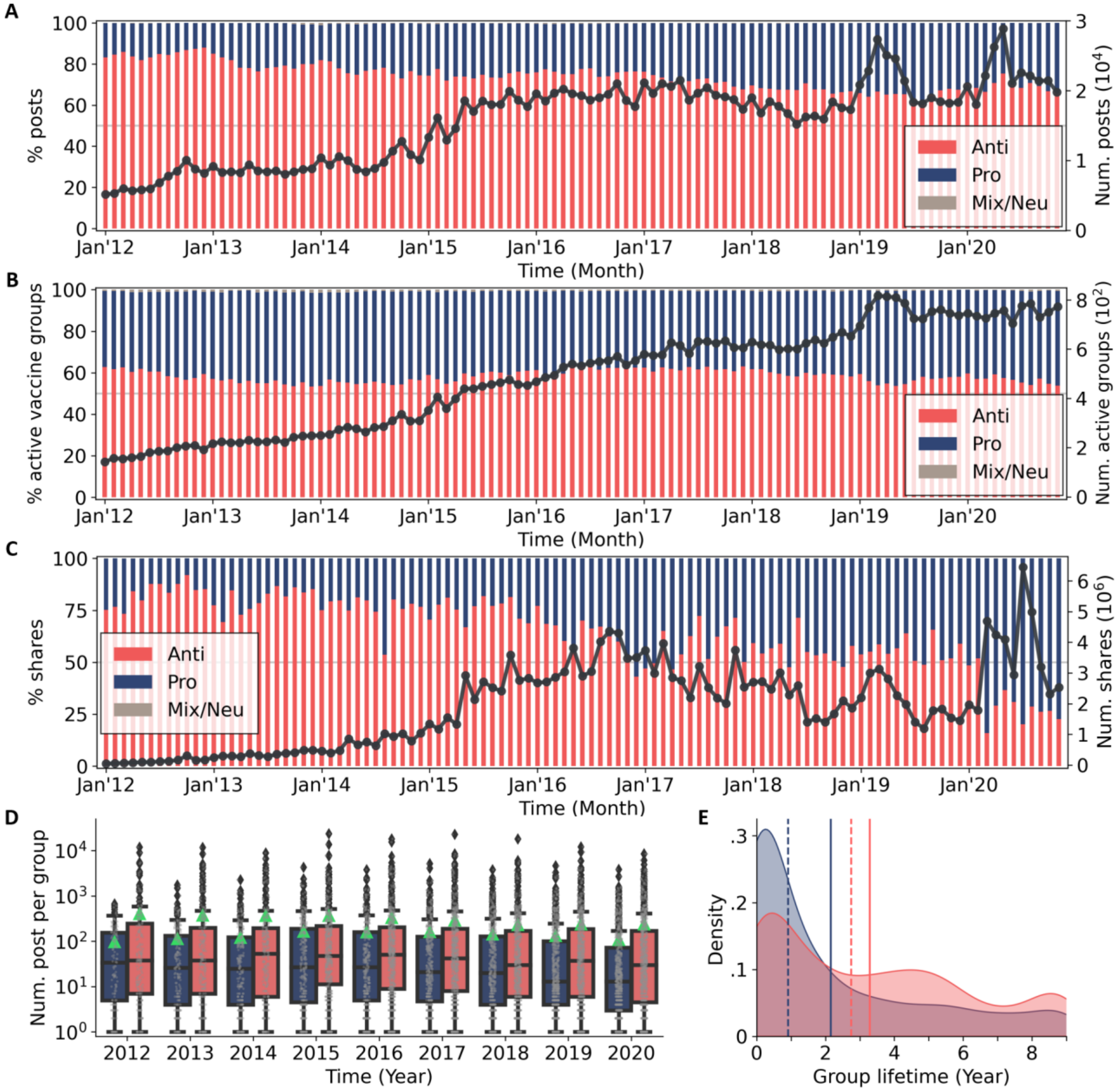

**Fig. 1. Prevalence and activity of vaccine groups.** **A.** Content created by vaccine groups. A stacked bar represents the proportions of Facebook posts created by pro-vaccine, anti-vaccine, and mixed/neutral groups in a given month (y-axis on the left). A dot indicates the total number of posts created by all vaccine groups in a given month (y-axis on the right). **B.** Active groups. A stacked bar represents the proportions of active pro-vaccine, anti-vaccine, and mixed/neutral groups in a given month (y-axis on the left). A dot indicates the number of all active groups in a given month (y-axis on the right). **C.** Sharing of vaccine content. A stacked bar represents the proportions of shares that pro-vaccine, anti-vaccine, and mixed/neutral groups received in total in a given month (y-axis on the left). A dot indicates the total number of shares in a given month (y-axis on the right). **D**. Activity level of groups. A box represents the distribution of the numbers of posts created by active pro- or anti-vaccine groups in a given year. A green triangle indicates an average. A gray dot represents an active group in a given year. **E**. Lifetime duration of vaccine groups. 1 year is equal to 365 days. The solid and dashed lines represent means and medians, respectively.

What information sources do anti-vaccine groups use to support their misleading narratives? To explore this question, we first examined the use of Facebook internal sources (i.e., information sources within the Facebook platform) and external sources (i.e., information sources outside Facebook) among anti- and pro-vaccine groups. The proportion of posts using Facebook internal sources has increased 7.0 times among pro-vaccine groups and 4.7 times among anti-vaccine groups between 2012 and 2020, while the use of external sources has steadily decreased (Fig. 2A). We also examined the proportions of video posts based on Facebook videos and YouTube videos, as an important case analysis comparing the use of Facebook internal sources and external sources. Consistent with the aforementioned patterns, the proportion of video posts using Facebook videos has increased dramatically between 2012 and 2020 among both anti- and pro-vaccine groups, in stark contrast to the substantial decline in the proportion of YouTube video posts (Fig. 2B). See S4 for details.

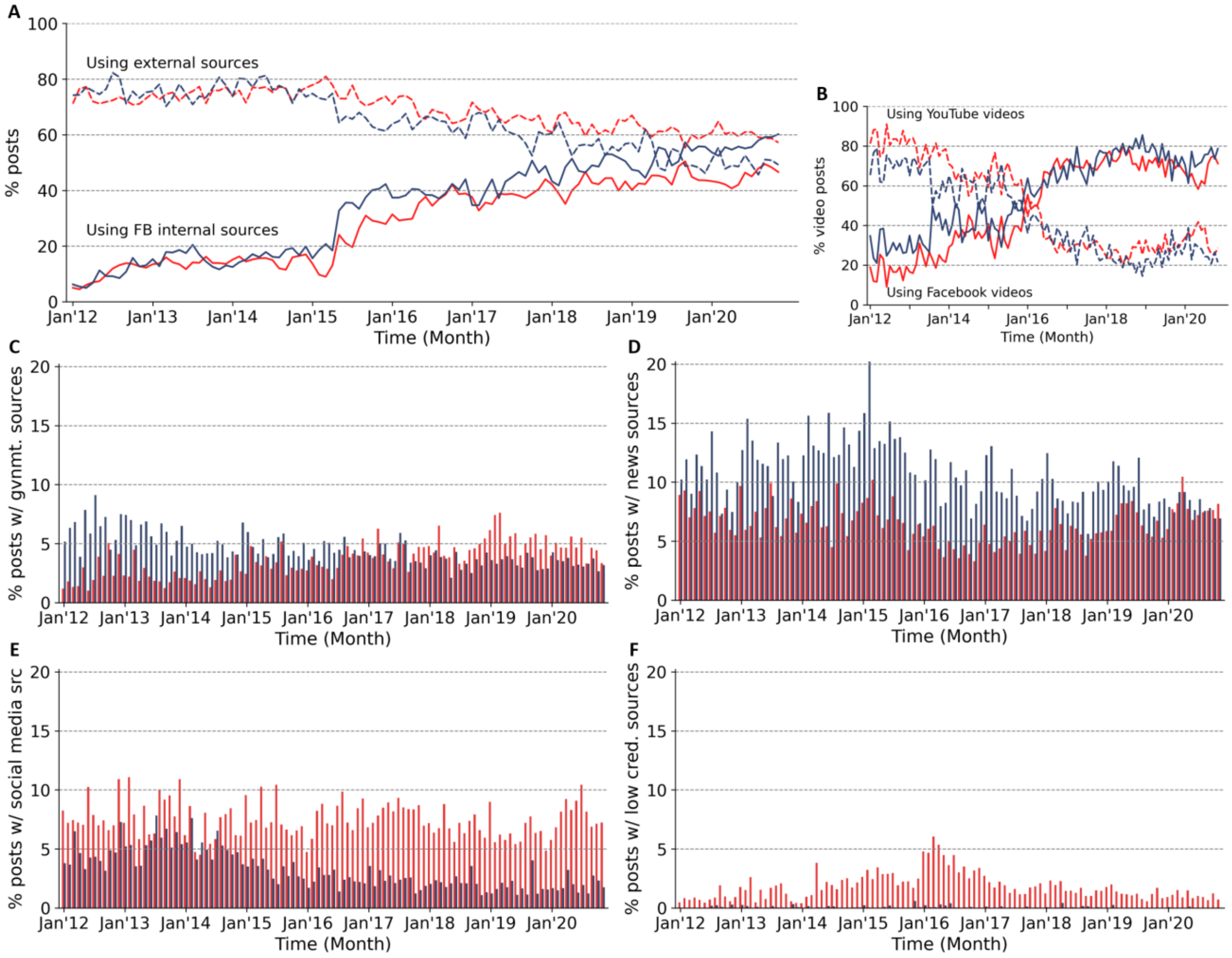

**Fig. 2. Information sources of anti- and pro-vaccine groups. A.** The proportions of posts using at least one Facebook internal source in an active group in a given month (solid lines) and the proportions of posts using at least one external source in an active group in a given month (dashed lines). Blue and red represent pro- and anti-vaccine, respectively. **B.** The proportion of video posts using Facebook videos (solid lines) and YouTube videos (dashed lines) in a vaccine group that generated at least one video post in a given year. **C.** The averaged proportions of posts using government sources in an active anti- and pro-vaccine group each month. Blue and red represent pro- and anti-vaccine, respectively. **D**. News sources. **E.** Social media sources. **F.** Low credibility sources.

We compared anti- and pro-vaccine groups in terms of their reliance on external sources known to supply misinformation (*15*, *17*) and other important types of external sources: government sources (*24*, *29*), social media sources (*27*, *28*), and news sources (*15*, *17*, *21*, *22*). As shown in Fig. 2F, the monthly proportion of posts using low credibility sources in an anti-vaccine group was generally limited with an average of 1.7% (*SD* = 0.9%), but it was still greater than the monthly average proportion in a pro-vaccine group, 0.1% (*SD* = 0.05%). Anti-vaccine groups used social media sources more frequently than pro-vaccine groups (Fig. 2E), and their reliance on news sources was lower than pro-vaccine groups (Fig. 2D). However, we identified significant recent increases in the use of legitimate sources among anti-vaccine groups, such as government (Fig. 2C) and news sources, while their use of low credibility sources has significantly decreased since 2016. See S4 for details.

We tested if posts using certain types of information sources attracted more or less engagement from Facebook users. As shown in Fig. 3A and B, posts using low credibility sources produced 2.4 times more shares than posts not using these sources, adjusting for covariates. On the other hand, posts using government sources received 52% fewer shares than posts not using these sources, and posts using social media sources attracted 55% fewer shares than posts without the sources. Notably, these associations were consistent for other measures of user engagement, comments and reactions, as shown in Fig. 3C and D. See S5 for details.

**A**

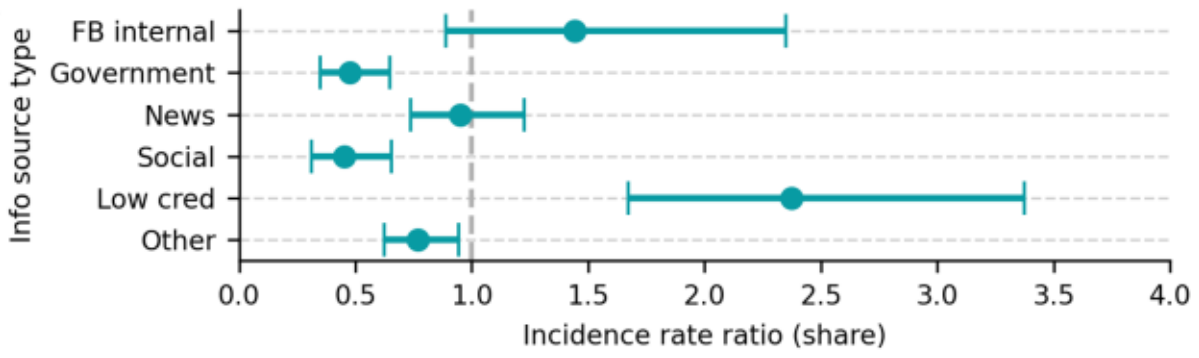

**B**

| | coef | IRR | 95% CI | std err | z | P > \|z\| |
|---|---|---|---|---|---|---|
| FB internal source | 0.3669 | 1.4433 | [0.8870, 2.3484] | 0.248 | 1.477 | 0.140 |
| Government source | -0.7424 | 0.4760 | [0.3491, 0.6490] | 0.158 | -4.694 | <0.001 |
| News source | -0.0518 | 0.9495 | [0.7356, 1.2255] | 0.130 | -0.398 | 0.691 |
| Social media source | -0.7948 | 0.4517 | [0.3114, 0.6552] | 0.190 | -4.188 | <0.001 |
| Low credibility source | 0.8649 | 2.3749 | [1.6723, 3.3726] | 0.179 | 4.833 | <0.001 |
| Other source | -0.2631 | 0.7687 | [0.6254, 0.9448] | 0.105 | -2.500 | 0.012 |
| Vaccine opposition | 0.4410 | 1.5542 | [1.1008, 2.1945] | 0.176 | 2.506 | 0.012 |
| Photo | 0.4158 | 1.5156 | [0.9419, 2.4389] | 0.243 | 1.713 | 0.087 |
| Video | 0.4960 | 1.6421 | [1.1686, 2.3076] | 0.174 | 2.857 | 0.004 |
| Subscriber | 1.989e-06 | 1.0000 | [1.0000, 1.0000] | 5.21e-07 | 3.819 | <0.001 |
| Month | 0.0108 | 1.0109 | [1.0059, 1.0158] | 0.003 | 4.303 | <0.001 |
| FB group (ref: FB page) | -3.1031 | 0.0449 | [0.0313, 0.0644] | 0.184 | -16.878 | <0.001 |
| Intercept | 2.7534 | 15.6964 | [7.5026,32.8390] | 0.377 | 7.311 | <0.001 |

**C**

**D**

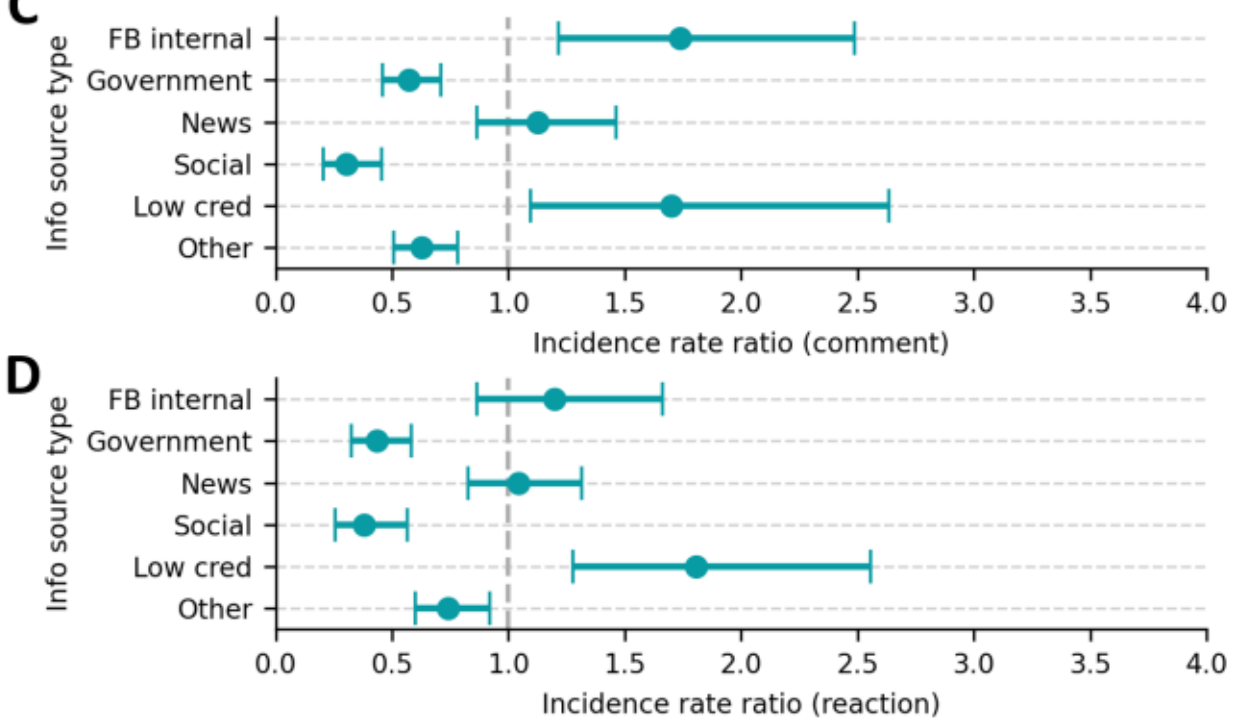

**Fig. 3. Models estimating associations between the use of information sources and user engagements. A.** Associations between the use of different source types in a post and the number of shares of the post. A dot and an error bar indicate the mean and the 95% confidence interval (CI) of an incidence rate ratio (IRR). Social, social media sources other than Facebook; Low cred, low credibility sources; Other, sources not classified into the first five types. *N* = 1,446,275. **B**. Results of a negative binomial model estimating the number of shares of a post as a function of variables shown in the leftmost column. Coef, regression coefficient; z, z score. All standard errors were clustered at the vaccine group level, and all models were estimated

with cluster-robust standard error at the vaccine group level. **C**. Associations between source types and comments. **D**. Associations between source types and reactions.

How do anti- and pro-vaccine groups interact within the broader ecosystem? First, we examined how many information sources were shared between anti- and pro-vaccine groups. For this purpose, exclusive sources for anti-vaccine [pro-vaccine] groups were defined as information sources used only by five or more vaccine groups with the same views in a given month. When averaged over the 107 months, exclusive sources accounted for 19.3% of all information sources used by an anti-vaccine group in a month (*SD* = 4.1%, Fig. 4A), while exclusive sources for pro-vaccine groups were only 3.1% of all sources that a pro-vaccine group used each month (*SD* = 1.5%, Fig. 4B) (*30*). See S6 for details. Second, we investigated if vaccine groups with different views relied on media sources emphasizing different values and beliefs. An anti-vaccine affinity score was calculated for each information source, which indicates the degree to which a source was used more widely by anti-vaccine groups than pro-vaccine groups (SM S7). We found that sources representing more conservative and far-right political views were more widely used by anti-vaccine groups, as the significant and positive correlation shown in Fig. 4C indicates. Lastly, we investigated the network of vaccine groups based on URL links between them. As the diagram in Fig. 4D shows, vaccine groups displayed insular patterns of interactions, forming two dense clusters separated from each other. Specifically, the cross-cutting connections between vaccine groups were only 5.3% (*SD* = 15.6%, *N* = 744) and 11.0% (*SD* = 22.7%, *N* = 410) of all connections that an anti- and a pro-vaccine group made with other groups, respectively, in the largest connected component (SM S8).

The current research studied the activities of online groups with different views on vaccines and the information ecosystem enabling the production of misleading anti-vaccine narratives. The results demonstrated that anti-vaccine groups had dominated Facebook in terms of the number of groups and the volume of content generated, which is aligned with previous reports based on smaller observations (*11*, *14*). The analysis showed that anti-vaccine groups relied more on low credibility sources than pro-vaccine groups did, but these sources were not widely used on both sides. Anti-vaccine groups also utilized social media sources more than pro-vaccine groups did, but their use of more legitimate sources, such as government sources and news sources, has also increased in recent years. Lastly, we discovered that the set of information sources used by anti-vaccine groups is, not only considerably segregated from those utilized by pro-vaccine groups, but also likely to represent more conservative and far-right viewpoints than pro-vaccine groups' sources.

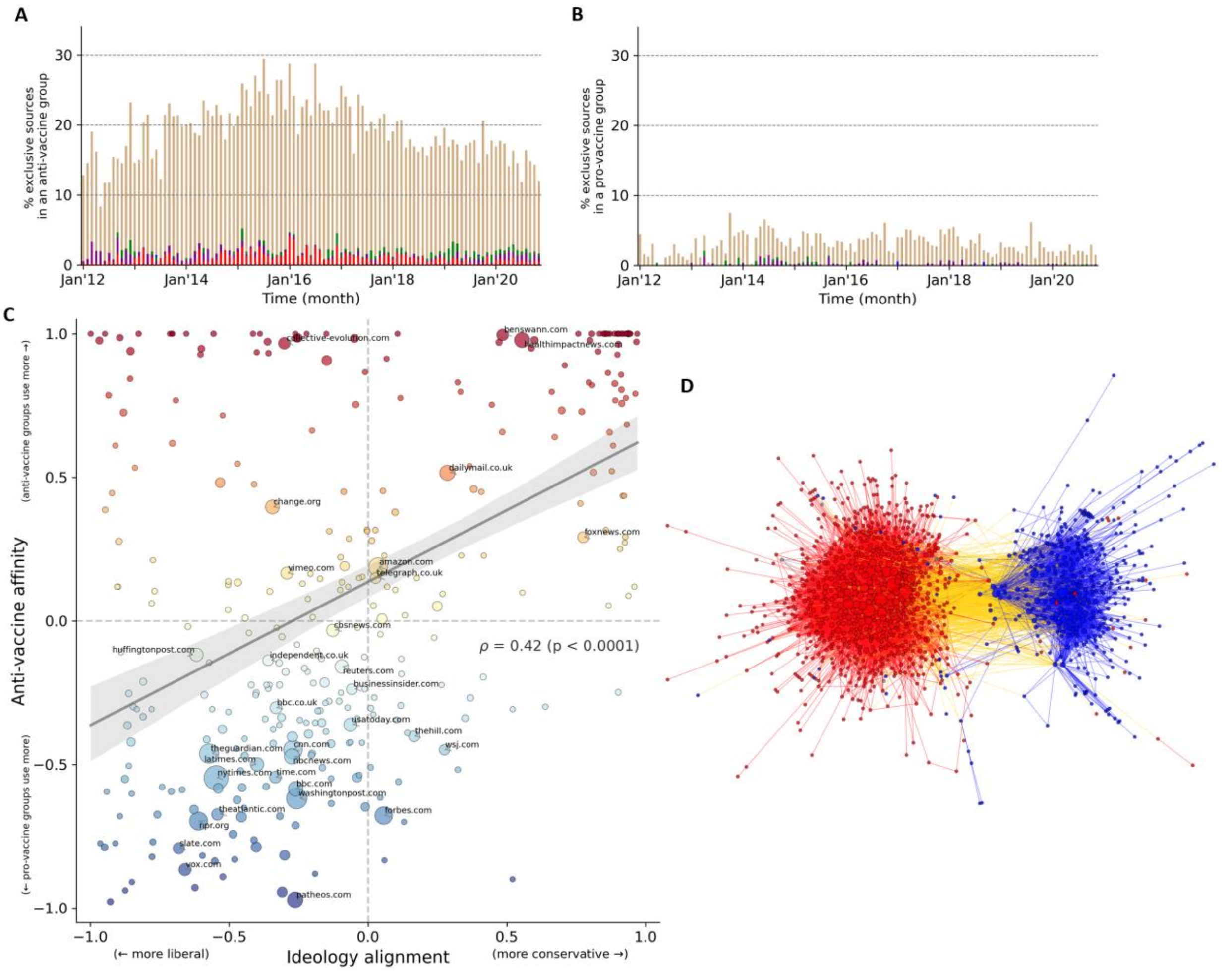

**Fig. 4. Anti- and pro-vaccine groups within the information ecosystem. A.** Exclusive sources for anti-vaccine groups. The height of each bar indicates the proportion of exclusive sources for anti-vaccine groups among all information sources in an anti-vaccine group, averaged across all anti-vaccine groups using at least one source in a given month. The red, blue, green, purple, and brown, portions of a bar represent the proportions of exclusive low credibility, social media, government, news, and other sources respectively, in an anti-vaccine group. **B.** Exclusive sources for pro-vaccine groups. **C.** Association between the ideological alignment and anti-vaccine affinity of sources. Each circle represents an information source. Ideology alignment scores were adopted from Bakshy et al. (21). The color of a circle depends on the source's anti-vaccine affinity. The size of a circle is proportional to the average popularity of a source. The solid line represents the linear regression line relating the two scores, and the shade represents the 95% CI of its slope. $\rho$ is the Spearman rank correlation between the two scores. $N$ = 327. See S7 for details. **D.** The vaccine group network within Facebook. The diagram shows the largest connected component including 1,059 vaccine groups. The red, blue, gray circles are anti-vaccine, pro-vaccine, and mixed/neutral groups, respectively. An edge indicates that two groups were connected via one or more URLs. Edge colors represent connections between anti-anti (red), anti-pro (yellow), and pro-pro(blue) vaccine groups, and connections involving mixed/neutral groups colored gray. The size of a circle is proportional to its degree.

This study is not without limitations. Although it was successful in identifying a large number of Facebook pages and groups discussing vaccines, data on accounts and content removed by Facebook, individual profiles, and private accounts are not available to researchers. Facebook argued that it would remove posts promoting false claims about COVID-19 vaccines (*31*), so the pervasiveness of anti-vaccine content before this alleged effort should have been more severe than what was identified in this research. The influence of algorithms and social media policies including content moderation also could not be considered due to the lack of available data and the limited collaboration between academia and the social media platform.

Despite the limitations, this study provides evidence that is urgently needed to understand and reduce misleading narratives about vaccines and vaccination on social media. First, hindering the processes of combining and merging Facebook internal resources to support anti-vaccine narratives, and disrupting Facebook's homegrown ecosystem enabling the production of disinformation should be a major target of intervention and policies aiming for vaccine disinformation reduction. (*8, 11*). Second, the intrusion of information from "fake news sources" is only one part of the problem. Both anti- and pro-vaccine groups use government, news, and social media sources more than low credibility sources, and anti-vaccine groups also seem to strive to be perceived as credible by using more and more government and legitimate news sources. Thus, although monitoring the use of suspicious sources could be a good way to identify users who are susceptible to misinformation (*15, 17*) and to detect incorrect information likely to go viral (*16*), strategies focusing only on these sources may not be sufficient in curbing false narratives. Lastly, anti-vaccine groups draw information from sources physically and politically distinguished from those of pro-vaccine groups, and the insular pattern of connections among online vaccine groups further limits information flows across groups with different views. Therefore, without considering the segregation between anti- and pro-vaccine groups on social media, public campaigns and interventions aiming to distribute scientific information may not be successful in reaching and influencing individuals with anti-vaccine beliefs.

**Acknowledgments:** The authors thank Nathaniel Persily, Rob Reich, Robert Hornik, and Sandra González-Bailón for their comments.

**Author contributions:**

Conceptualization: SK

Methodology: SK

Investigation: SK, KK

Visualization: SK, KK

Writing – original draft: SK, KK

Writing – review & editing: SK, KK

**Competing interests:** Authors declare that they have no competing interests.

**Data and materials availability:** All data, code, and materials used in the analysis are available in the supplementary materials.

**Supplementary Materials**

Figs. S1 to S14

Tables S1 to S14

References (1–48)

# Supplementary Material for "The Information Ecosystem of Online Groups with Anti- and Pro-vaccine Views on Facebook"

## TABLE OF CONTENTS

# S1 Definition and Terminology

## S1.1 Vaccine Group and Vaccine Content

In the current study, a “group” refers to a community of Facebook users who created, consumed, or engaged with the content of a Facebook page or Facebook group. A “vaccine group” refers to a group discussing vaccine-related issues. When referring to a webpage within Facebook that is either a Facebook page and a Facebook group, we use the term a “Facebook page/group” (without omitting “Facebook”) or, simply put, an “account.” Compared with “groups,” “Facebook pages/groups” or “accounts” were used to highlight the online content that communities generated or the online spaces in which communities communicated. “Vaccine content” or “vaccine posts” refer to Facebook posts created by vaccine groups.

A pro-vaccine group refers to a vaccine group that advocates safety, effectiveness, and benefits of vaccines and vaccination; encourages perception and norms favorable to vaccines; refutes claims about the danger of and suspicions about vaccines and criticizes people supporting these claims; and/or provides information about services, policies, or campaigns for vaccines and related practices. An anti-vaccine group, on the other hand, refers to a vaccine group that emphasizes danger and negative consequences of vaccines and vaccination; promotes information about vaccine exemptions and choice; highlights the benefits of other medical or dietary practices as alternatives; presents claims based on conspiracy theories and unsubstantiated information; and/or criticizes pro-vaccine content and people who promote it.

In this research, we distinguish between misleading narratives and misinformation. Misleading narratives refer to claims and stories promoted by anti-vaccine groups. Misinformation, on the other hand, refers to information drawn from low credibility information sources. The inclusion of misinformation in a Facebook post is determined by the existence of URLs linked to low credibility information sources, not by the view of the vaccine group producing the post. Thus, the inclusion of misinformation in a post does not necessarily indicate that the post was produced by an anti-vaccine group; misinformation can be included in content produced by not only anti-vaccine groups but also pro-vaccine groups.

An active group refers to a group that created at least one post in a given period of time. For example, “active groups in January 2015” refers to vaccine groups that generated at least one post in that month, and “active groups in 2017” refers to vaccine groups that generated at least one post in that year. In the same sense, an active anti-vaccine group [an active pro-vaccine group] indicates an anti-vaccine group [a pro-vaccine group] that created at least one post in a given period of time.

## S1.2 Information Source

This research explored the information ecosystem that enabled vaccine groups by analyzing URLs included in the posts they generated and identifying information sources used by them. Information sources used in a Facebook post can be identified by analyzing URLs (Unified Resource Locators) included in the post. Each URL (e.g., https://www.cnn.com/health) indicates a source from which a piece of information was drawn (e.g., cnn.com). Multiple URLs can be included in a post, and data retrieved from CrowdTangle provide information on all URLs in a post. For example, if a post in Facebook page A includes a hyperlink to a CDC webpage, a video uploaded on Facebook, and an URL to a Washington Post article, the data on the A’s post will include the URL of the CDC webpage, the URL of the video on Facebook, and the URL of the Washington Post article. In this case, the post in Facebook Group A is drawing information from three different information sources: cdc.gov, facebook.com, and washingtonpost.com.

# S2 Data Collection and Valence Coding

## S2.1 Overview

CrowdTangle (CT) is "a public insights tool owned and operated by Facebook." It is one of the very few tools that help researchers and journalists monitor a large number of public Facebook pages and groups. CT maintains data on public Facebook accounts that are selected by CT based on their popularity or are requested by CT users (*1*, *2*). Based on these features of CT, we designed an iterative data collection method combining data retrieval through CT and the identification of new Facebook pages/groups. This method, which was inspired by the snowball sampling methods, enabled us to gather a large number of Facebook pages/groups that satisfy our search criteria beyond the limited number of accounts originally offered by CT. The data collection was conducted between December 13th and December 25th, 2020.

Although, without access to the company's internal knowledge base about the Facebook website and database architecture, it is currently impossible for independent researchers to identify and access a complete set of Facebook pages/groups that meet certain search criteria, we believe that the data collection method of the present study has made a meaningful contribution in expanding the scope of data that can be retrieved from Facebook.

This study was approved by the Institutional Review Board of [removed for the blind review process].

## S2.2 Data Collection Procedure

Data collection consisted of two stages. In the first stage (Stage 1), iterative keyword-based searching was conducted. Stage 1 consisted of the four rounds of data collection. In the second stage (Stage 2), we conducted network-based searching based on the results of the first stage. Stage 2 aimed to capture additional Facebook pages/groups that were connected with accounts found in Stage 1 but could not be captured by a keyword-based searching strategy.

We considered the technical characteristics of CT when devising the data collection strategy for the current study. Using CT is, as of now, the only way available for external researchers who aim to search for specific Facebook pages/groups and to download content created by these accounts. However, after an extensive examination of the CT's search function, we identified the following functional irregularities. (i) The search results were not exhaustive. The search function did not provide all matching results in one search attempt. We discovered that more results could be obtained by accompanying related keywords with a keyword of interest. For example, the searching results of a keyword sequence "vaccin danger" did not necessarily constitute a subset of the searching results of "vaccin." (ii) The search results were not exact. The search results might include Facebook pages/groups with titles containing possible variations of keywords. We speculated that it was because of Facebook's consideration of potential user typos. For example, the search results of "vaccin" might contain a Facebook page title with other words with similar spelling, such as "vacation" or "vacuum." We contacted CrowdTangle for these issues, and they responded that CT's search function utilized Facebook's search engine and that CT had only a limited ability to investigate, correct, or publicize technical details of Facebook's proprietary search function.

#### S2.2.1 Stage 1: Iterative Keyword-based Searching

Stage 1 conducted keyword-based searching and additional filtering to identify Facebook pages/groups discussing vaccine-related issues. Posts created by the resulting Facebook accounts were then retrieved through CT. The design of Stage 1 considers the aforementioned functional irregularities and aims to overcome them.

Round 1 followed the following steps. First, utilizing the page/group search function that CT's web interface provides, we searched for Facebook pages/groups discussing vaccine-related topics using various combinations of "core" and "supporting" keywords. To offset the incompleteness of CT's search results, multiple searches were conducted using multiple keywords. Table S1 shows all combinations of core and supporting keywords used in this step. As shown in the table, the two core keywords are "vaccin" and "vax," and supporting keywords are a set of keywords that are commonly used in the discussion of vaccine-related issues, such as "prevent" and "danger." Second, after retrieving from CT a full list of the Facebook pages/groups that corresponds the keyword combinations, we conducted additional filtering to select accounts containing one or more of the two core keywords in their titles. This step was to overcome the inexactness of CT's search results by filtering out Facebook pages/groups that did not contain any core keywords. Third, using CT's Post Search API, we downloaded English posts created between January 1st, 2012 and November 30th, 2020 by the Facebook pages/groups that passed the first and second steps.

Round 2 started by analyzing all Facebook internal URLs included in the post data downloaded in Round 1 and identifying new Facebook accounts that were not included in the list of Facebook pages/groups identified in Round 1. We requested CT to update its database if it had not been tracking these newly found accounts. Among these new accounts, we selected Facebook pages/groups including one or more core keywords in their titles and retrieved all English posts created by the selected accounts from January 1st, 2012 to November 30th, 2020 using the Post Search API.

Similarly, Round 3 started by analyzing all Facebook internal URLs included in the post data downloaded in Round 2 and identifying new Facebook accounts that were not included in the list of Facebook pages/groups identified in the previous rounds. We requested CT to update its database if it had not been tracking the newly found accounts. Among the new accounts identified in Round 3, we selected Facebook pages/groups including one or more core keywords in their titles. All English posts created by the selected accounts from January 1st, 2012 to November 30th, 2020 were retrieved using the Post Search API.

Round 4 also started by analyzing all Facebook internal URLs included in the post data downloaded in Round 3 and identifying new Facebook accounts that were not included in the list of Facebook pages/groups identified in the previous rounds. We requested CT to update its database if it had not been tracking these newly found accounts. Among the new accounts identified in Round 4, we selected Facebook pages/groups including one or more core keywords in their titles. All English posts created by the selected accounts from January 1st, 2012 to November 30th, 2020 were retrieved using the Post Search API. Because only one accounts were newly identified in this round, we stopped Stage 1 in this round.

Lastly, we merged all post data retrieved in Round 1 to 4. The resulting data contained posts from 2,079 Facebook pages/groups.

**Table S1.** Keyword Combinations for Stage 1

| | | |
|---|---|---|
| "pro vaccin vax" | "vaccin vax death" | "vaccin vax cause" |
| "vaccin vax good" | "vaccin vax choice" | "vaccin vax induce" |
| "vaccin vax safe" | "vaccin vax consent" | "vaccin vax disease" |
| "vaccin vax save" | "vaccin vax against" | "vaccin vax alliance" |
| "vaccin vax protect" | "vaccin vax stop" | "vaccin vax coalition" |
| "vaccin vax prevent" | "vaccin vax concern" | "vaccin vax movement" |
| "vaccin vax benefit" | "vaccin vax victim" | "vaccin vax action" |
| "vaccin vax immun" | "vaccin vax fact" | "vaccin vax" |
| "vaccin vax advoca" | "vaccin vax truth" | "vaccine" |
| "vaccin vax support" | "vaccin vax real" | "vaccinate" |
| "vaccin vax science" | "vaccin vax hoax" | "vaccinated" |
| "vaccin vax access" | "vaccin vax myth" | "vaccination" |
| "anti vaccin vax" | "vaccin vax aware" | "vaccinating" |
| "vaccin vax danger" | "vaccin vax inform" | "vaxxed" |
| "vaccin vax risk" | "vaccin vax expert" | "vaxxer" |
| "vaccin vax injury" | "vaccin vax educat" | "vaxxing" |
| "vaccin vax damage" | "vaccin vax know" | |
| "vaccin vax kill" | "vaccin vax mandat" | |

### S2.2.2 Stage 2: Network-based Searching

To discover more Facebook pages/groups discussing vaccine-related issues, Stage 2 conducted the following tasks: (a) Analyzing Facebook internal URLs included in the posts found in Stage 1, (b) identifying candidate accounts, i.e., Facebook pages/groups connected with the Stage 1 accounts via URLs, (c) among the candidates, requesting CT to obtain data on newly found accounts that did not exist in its database, (d) selecting accounts satisfying at least one of the two conditions. The two conditions were: (1) a Facebook page/group was referenced by at least 5 different groups found in Stage 1, (2) a Facebook page/group contains at least one of the following keywords in its title: “vac”, “vax”, “immun”, and “shot”. These processes aimed to collect additional accounts that discuss vaccination but had not been captured in Stage 1 because their titles did not meet Stage 1’s keyword-searching criteria. (Note that Stage 2’s keyword rule was loosened from that of Stage 1, considering that the candidate accounts of Stage 2 were already being referenced by other Facebook pages/groups that satisfied the criteria of Stage 1.) 391 Facebook pages/groups were newly discovered in Stage 2, and we retrieved their English posts produced from January 1st, 2012 to November 30th, 2020.

We merged all data retrieved in Stage 1 and Stage 2. The resulting data included 2,462 accounts (2,315 Facebook pages, 147 Facebook groups).

## S2.3 Evaluation of Group Valence

We examined the view of each Facebook page and group on vaccines and vaccination. For this purpose, each Facebook page or group was evaluated by two independent coders and classified into one of the four categories: "pro-vaccine," "anti-vaccine," "mixed/neutral," and "not applicable." The classification of pro-vaccine and anti-vaccine groups was based on the definitions provided in S1. A Facebook page/group was labeled as mixed/neutral if it presented mixed viewpoints, or views that were related to vaccines but neither pro- nor anti-vaccine. The "not applicable" category included pages/groups that did not discuss vaccine-related issues (e.g., musicians and metaphorical use of the word, vaccine), were mostly about animal or non-human vaccines, did not contain any content written in English, was not accessible, or did not contain two or more posts about vaccines and vaccination.

Two subject-matter experts coded each Facebook page/group. Before starting the main coding task, all coders reviewed the same set of 100 randomly selected pages/groups independently, following the coding procedure described below. All coders then shared and compared their coding results and discussed and resolved discrepancies in their understanding. Coders then coded the rest of the pages/groups. Each page/group was coded by two independent coders.

The coding procedure was as follows. First, each coder checked the description of each page/group in its "About" tab on Facebook and attempted to determine its view based on the information provided in the description. If the description did not provide sufficient information, coders reviewed up to 25 most recent postings in a page/group, including posts, photos, status updates, and added events. A coding decision was made based on at least two postings related to vaccines or vaccination.

The coding results showed that the evaluations of two coders were identical for 2,373 of 2,462 Facebook pages/groups. Krippendorff's alpha was .936. For the 89 pages/groups that two coders did not agree upon, two coders made final decisions together after discussion. The coding results also indicated that that the collected data included 134 “not applicable” Facebook pages/groups, and we removed these accounts from the dataset. Consequently, the final dataset contained 1,682,205 million posts created by 2,328 vaccine groups (1,206 pro-vaccine, 1,083 anti-vaccine, and 39 mixed/neutral groups; 2185 Facebook pages and 143 Facebook groups) from January 1st, 2012 to November 30th, 2020.

## S2.4 Comparison with the Data Collected by Previous Studies

We have reviewed prior studies that analyzed communities discussing vaccines and vaccination on social media (*3–16*). Here, we compare the current study with four of the most comprehensive previous studies on Facebook vaccine communities, from the perspectives of data scales, collection methods, and analytical approaches.

First, Johnson et al. (*15*) manually identified 441 publicly accessible Facebook pages with pro- and anti-vaccine views in 2019. Specifically, they first chose multiple “seed” pages about vaccination and identified pages liked by the seed pages’ administrators. The researchers then visited these “liked pages” and identified other pages liked by the liked pages’ administrators. By repeating this process, the researchers found 317 “anti-vaccination,” 124 “pro-vaccination,” and 885 “undecided” pages (*15*). (Based on a similar sampling method, Broniatowski et al. (*14*) identified 204 “anti-vaccination” Facebook pages in 2019 and downloaded 288,175 posts created by them through CT.) Thus, the current study is more extensive than Johnson et al., considering the number of online communities identified and the duration of data collected for research. The data collected in the current study also include both Facebook groups and pages, while Johnson et al. targeted only the latter. Furthermore, we investigated the content of these Facebook pages and groups, which was only qualitatively discussed in Johnson et al. In addition, the data collection method of the present study helps identify a set of Facebook pages and groups that are more closely related to the vaccines and vaccination issues, while the method following Facebook accounts liked by page administrators might overemphasize the preference of a small number of individuals.

Second, Schmidt et al. (*6*) identified 243 Facebook pages discussing vaccines and vaccination between January 2010 and May 2017, including 145 pro-vaccination or 98 anti-vaccination Facebook pages. They focused on the volume of content and the number of user engagements over time. Cinelli et al. (*16*) re-analyzed the data collected in Schmidt et al. and reported that Facebook users with similar views on vaccination were more likely to interact with each other.

# S3 Analysis: The Activities of Pro- and Anti-vaccine Groups

## S3.1 The Volume of Content

Table S2 shows the number of posts generated by all vaccine groups each month and the proportion of posts from anti-vaccine, pro-vaccine, and mixed/neutral groups. The 107-month average of the total number of posts was 15721.5 (*SD* = 5703.4, Median = 18005). The 107-month average proportions of posts created by anti-vaccine, pro-vaccine, and mixed/neutral groups were 74.1% (*SD* = 6.2%, Median = 74.3%), 25.6% (*SD* = 6.2%, Median = 25.2%), and 0.2% (*SD* = 0.2%, Median = 0.2%, *N* = 107), respectively.

As shown in Fig 1A, the total number of posts produced by all vaccine groups had increased steadily with noticeable surges that coincide with three major infectious disease outbreaks in the U.S: (a) the 2015 U.S. measles outbreak that peaked in January 2015 (*17*), (b) the 2019 U.S. measles outbreak that peaked in March 2019 (*18*), and (c) the early stage of the COVID-19 pandemic. (The U.S. National Emergency concerning COVID-19 was declared in March 2020 (*19*).)

To examine if the proportion of posts generated by anti-vaccine groups was mostly determined by a few Facebook pages/groups that published a large number of posts, we also calculated the proportion excluding the top 10 vaccine groups in terms of the total number of posts generated. As shown in Fig. S1A, the calculation showed that the majority of vaccine posts were produced by anti-vaccine groups even when the top 10 vaccine groups were excluded. Specifically, the 107-month average proportions of posts created by anti-vaccine, pro-vaccine, and mixed/neutral groups were 64.7% (*SD* = 5.2%, Median = 64.0%), 35.0% (*SD* = 5.3%, Median = 35.6%), and 0.4% (*SD* = 0.3%, Median = 0.3%, *N* = 107), respectively. When the top 20 vaccine groups with the most posts were removed from the dataset, the posts created by anti-vaccine groups were still dominant (the 107-month average = 60.5%, *SD* = 4.3%, Median = 60.2%), as shown in Fig. S1B.

**Table S2**. Number of posts, active groups, and shares by month

| | Posts | | | | Active groups | | | | Shares | | | |
|---|---|---|---|---|---|---|---|---|---|---|---|---|
| Month | Total num | % Anti | % Pro | % Mix | Total num | % Anti | % Pro | % Mix | Total num | % Anti | % Pro | % Mix |
| 01/2012 | 5125 | 83.3 | 16.6 | 0.2 | 142 | 62.7 | 36.6 | 0.7 | 48077 | 75.3 | 24.7 | 0.0 |
| 02/2012 | 5285 | 84.6 | 15.2 | 0.2 | 157 | 61.8 | 37.6 | 0.6 | 55084 | 76.7 | 23.3 | 0.0 |
| 03/2012 | 5989 | 85.9 | 14.0 | 0.1 | 155 | 62.6 | 36.8 | 0.6 | 68149 | 73.3 | 26.7 | 0.0 |
| 04/2012 | 5664 | 83.6 | 16.1 | 0.2 | 159 | 60.4 | 38.4 | 1.3 | 75162 | 84.1 | 15.9 | 0.0 |
| 05/2012 | 5819 | 82.0 | 17.7 | 0.3 | 165 | 61.8 | 37.0 | 1.2 | 97770 | 79.8 | 20.2 | 0.0 |
| 06/2012 | 5946 | 83.2 | 16.5 | 0.3 | 180 | 60.6 | 38.3 | 1.1 | 92443 | 87.7 | 12.3 | 0.0 |
| 07/2012 | 6822 | 84.9 | 15.0 | 0.1 | 185 | 60.5 | 38.4 | 1.1 | 121877 | 87.7 | 12.3 | 0.0 |
| 08/2012 | 7760 | 84.4 | 15.4 | 0.2 | 187 | 58.3 | 40.6 | 1.1 | 130619 | 83.5 | 16.5 | 0.0 |
| 09/2012 | 8484 | 85.7 | 14.1 | 0.2 | 201 | 57.7 | 41.3 | 1.0 | 162594 | 87.7 | 12.3 | 0.0 |
| 10/2012 | 10020 | 86.8 | 13.2 | 0.0 | 207 | 56.5 | 42.5 | 1.0 | 306125 | 92.0 | 8.0 | 0.0 |
| 11/2012 | 8766 | 87.4 | 12.5 | 0.1 | 209 | 57.4 | 42.1 | 0.5 | 160503 | 84.8 | 15.2 | 0.0 |
| 12/2012 | 8178 | 87.9 | 11.9 | 0.1 | 192 | 59.4 | 39.6 | 1.0 | 169136 | 85.2 | 14.8 | 0.0 |
| 01/2013 | 9158 | 85.0 | 14.9 | 0.2 | 216 | 56.9 | 42.1 | 0.9 | 249192 | 77.4 | 22.6 | 0.0 |
| 02/2013 | 8279 | 83.2 | 16.6 | 0.1 | 224 | 56.3 | 42.4 | 1.3 | 300533 | 69.3 | 30.6 | 0.0 |
| 03/2013 | 8386 | 81.9 | 18.1 | 0.1 | 220 | 56.8 | 42.7 | 0.5 | 292544 | 84.6 | 15.4 | 0.0 |
| 04/2013 | 8264 | 78.2 | 21.7 | 0.1 | 220 | 55.0 | 44.5 | 0.5 | 275467 | 73.0 | 27.0 | 0.0 |
| 05/2013 | 9390 | 78.0 | 21.9 | 0.1 | 230 | 56.1 | 43.5 | 0.4 | 375414 | 75.6 | 24.4 | 0.0 |
| 06/2013 | 8530 | 76.4 | 23.4 | 0.2 | 224 | 54.9 | 43.8 | 1.3 | 312671 | 78.5 | 21.5 | 0.0 |
| 07/2013 | 8388 | 78.0 | 21.9 | 0.1 | 224 | 54.9 | 44.2 | 0.9 | 275942 | 83.0 | 17.0 | 0.0 |
| 08/2013 | 8516 | 78.5 | 21.4 | 0.2 | 231 | 56.3 | 42.9 | 0.9 | 353703 | 86.7 | 13.3 | 0.0 |
| 09/2013 | 8024 | 79.2 | 20.6 | 0.2 | 221 | 54.3 | 44.3 | 1.4 | 388157 | 81.8 | 18.2 | 0.0 |
| 10/2013 | 8399 | 77.7 | 21.7 | 0.5 | 243 | 53.5 | 45.3 | 1.2 | 403260 | 86.1 | 13.9 | 0.0 |
| 11/2013 | 8727 | 80.0 | 19.2 | 0.8 | 247 | 55.5 | 43.7 | 0.8 | 484771 | 83.6 | 16.4 | 0.0 |
| 12/2013 | 8823 | 80.0 | 19.2 | 0.7 | 249 | 53.4 | 45.0 | 1.6 | 487774 | 85.2 | 14.8 | 0.0 |
| 01/2014 | 10383 | 81.9 | 16.9 | 1.2 | 250 | 53.6 | 44.8 | 1.6 | 453513 | 75.2 | 24.8 | 0.0 |
| 02/2014 | 9322 | 81.4 | 18.2 | 0.4 | 254 | 56.7 | 42.1 | 1.2 | 398747 | 79.3 | 20.7 | 0.0 |
| 03/2014 | 10571 | 77.8 | 21.7 | 0.5 | 274 | 55.5 | 43.1 | 1.5 | 476616 | 79.7 | 20.3 | 0.0 |
| 04/2014 | 9979 | 75.5 | 23.9 | 0.5 | 284 | 55.3 | 43.3 | 1.4 | 855542 | 75.0 | 25.0 | 0.0 |
| 05/2014 | 8730 | 74.8 | 24.5 | 0.7 | 277 | 54.5 | 44.4 | 1.1 | 658972 | 81.2 | 18.8 | 0.0 |
| 06/2014 | 8398 | 76.7 | 22.8 | 0.5 | 264 | 54.9 | 44.3 | 0.8 | 756047 | 79.8 | 20.2 | 0.0 |
| 07/2014 | 8872 | 77.3 | 22.3 | 0.3 | 282 | 55.7 | 43.6 | 0.7 | 633659 | 74.4 | 25.6 | 0.0 |
| 08/2014 | 9730 | 78.1 | 21.6 | 0.3 | 286 | 54.9 | 44.4 | 0.7 | 1018336 | 53.7 | 46.3 | 0.0 |
| 09/2014 | 11356 | 75.3 | 24.4 | 0.3 | 309 | 54.0 | 45.0 | 1.0 | 927603 | 80.3 | 19.7 | 0.0 |
| 10/2014 | 12744 | 73.0 | 26.9 | 0.2 | 335 | 54.3 | 44.8 | 0.9 | 1026374 | 75.7 | 24.3 | 0.0 |
| 11/2014 | 10808 | 76.6 | 23.2 | 0.2 | 308 | 56.8 | 42.5 | 0.6 | 781481 | 78.2 | 21.8 | 0.0 |
| 12/2014 | 10072 | 74.4 | 25.1 | 0.5 | 310 | 56.5 | 42.6 | 1.0 | 1037039 | 76.7 | 23.3 | 0.0 |
| 01/2015 | 13337 | 74.3 | 25.2 | 0.5 | 351 | 59.0 | 40.5 | 0.6 | 1340063 | 70.6 | 29.4 | 0.0 |
| 02/2015 | 16132 | 77.5 | 21.9 | 0.6 | 406 | 56.9 | 41.9 | 1.2 | 1173656 | 77.7 | 22.3 | 0.0 |
| 03/2015 | 12933 | 72.0 | 27.9 | 0.1 | 359 | 54.6 | 44.8 | 0.6 | 1543655 | 81.1 | 18.9 | 0.0 |
| 04/2015 | 14573 | 73.9 | 25.8 | 0.3 | 399 | 55.9 | 43.4 | 0.8 | 1326061 | 75.5 | 24.5 | 0.0 |
| 05/2015 | 18563 | 74.0 | 25.8 | 0.2 | 439 | 59.7 | 39.6 | 0.7 | 2916403 | 66.8 | 33.2 | 0.0 |
| 06/2015 | 17050 | 73.0 | 26.8 | 0.2 | 440 | 58.2 | 40.7 | 1.1 | 2138934 | 77.0 | 23.0 | 0.0 |
| 07/2015 | 18510 | 74.9 | 25.0 | 0.2 | 449 | 59.9 | 39.4 | 0.7 | 2712394 | 81.8 | 18.2 | 0.0 |
| 08/2015 | 18005 | 73.4 | 26.3 | 0.3 | 457 | 60.6 | 38.5 | 0.9 | 2520747 | 77.1 | 22.9 | 0.0 |
| 09/2015 | 18017 | 73.3 | 26.4 | 0.3 | 464 | 60.1 | 39.4 | 0.4 | 2408841 | 78.3 | 21.7 | 0.0 |
| 10/2015 | 19936 | 75.5 | 24.1 | 0.3 | 475 | 58.7 | 40.6 | 0.6 | 3587137 | 81.5 | 18.5 | 0.0 |
| 11/2015 | 18634 | 76.2 | 23.5 | 0.3 | 457 | 60.6 | 38.7 | 0.7 | 2762723 | 71.0 | 29.0 | 0.0 |
| 12/2015 | 17740 | 75.5 | 24.2 | 0.3 | 453 | 60.9 | 38.4 | 0.7 | 2831317 | 68.8 | 31.2 | 0.0 |
| 01/2016 | 19589 | 75.9 | 23.7 | 0.4 | 468 | 61.3 | 38.0 | 0.6 | 2680810 | 77.2 | 22.8 | 0.0 |
| 02/2016 | 18500 | 77.4 | 22.1 | 0.5 | 485 | 58.8 | 40.2 | 1.0 | 2708973 | 68.4 | 31.6 | 0.0 |
| 03/2016 | 19638 | 76.3 | 23.2 | 0.6 | 494 | 60.1 | 39.1 | 0.8 | 2858237 | 67.7 | 32.3 | 0.0 |
| 04/2016 | 20242 | 75.1 | 24.5 | 0.4 | 528 | 59.5 | 40.0 | 0.6 | 3041838 | 60.2 | 39.8 | 0.0 |
| 05/2016 | 19570 | 75.1 | 24.6 | 0.3 | 540 | 63.9 | 35.6 | 0.6 | 3818994 | 54.6 | 45.4 | 0.0 |
| 06/2016 | 19315 | 77.5 | 22.0 | 0.4 | 532 | 64.1 | 35.0 | 0.9 | 2900978 | 70.1 | 29.9 | 0.0 |
| 07/2016 | 18658 | 77.9 | 21.7 | 0.4 | 544 | 64.0 | 35.3 | 0.7 | 3049748 | 66.3 | 33.7 | 0.0 |
| 08/2016 | 18999 | 73.7 | 25.9 | 0.4 | 550 | 61.5 | 37.8 | 0.7 | 4019756 | 67.1 | 32.9 | 0.0 |
| 09/2016 | 19488 | 74.0 | 25.7 | 0.3 | 554 | 62.3 | 37.0 | 0.7 | 4357140 | 62.4 | 37.6 | 0.0 |
| 10/2016 | 21005 | 75.9 | 23.8 | 0.3 | 571 | 62.3 | 37.0 | 0.7 | 4304908 | 59.9 | 40.1 | 0.0 |
| 11/2016 | 18585 | 76.2 | 23.5 | 0.3 | 536 | 62.5 | 36.8 | 0.7 | 3476339 | 53.6 | 46.4 | 0.0 |
| 12/2016 | 17747 | 76.4 | 23.3 | 0.3 | 553 | 62.4 | 37.1 | 0.5 | 3504012 | 43.1 | 56.9 | 0.0 |
| 01/2017 | 21138 | 76.2 | 23.6 | 0.2 | 580 | 62.4 | 37.1 | 0.5 | 3737598 | 47.1 | 52.9 | 0.0 |
| 02/2017 | 19613 | 74.6 | 25.2 | 0.2 | 575 | 60.5 | 39.0 | 0.5 | 2987266 | 49.5 | 50.5 | 0.0 |
| 03/2017 | 21011 | 73.4 | 26.4 | 0.2 | 577 | 62.6 | 37.1 | 0.3 | 3962861 | 65.1 | 34.9 | 0.0 |
| 04/2017 | 20617 | 71.5 | 28.3 | 0.2 | 626 | 61.8 | 37.5 | 0.6 | 2851254 | 46.6 | 53.4 | 0.0 |
| 05/2017 | 21478 | 72.5 | 27.3 | 0.2 | 616 | 61.2 | 38.1 | 0.6 | 2748132 | 53.7 | 46.3 | 0.0 |
| 06/2017 | 18632 | 71.6 | 28.2 | 0.2 | 582 | 62.7 | 36.8 | 0.5 | 2198230 | 62.6 | 37.4 | 0.0 |
| 07/2017 | 19643 | 72.7 | 27.0 | 0.2 | 632 | 61.6 | 37.7 | 0.8 | 3221800 | 72.3 | 27.7 | 0.0 |
| 08/2017 | 20454 | 73.0 | 26.7 | 0.3 | 631 | 62.9 | 36.5 | 0.6 | 2522100 | 51.6 | 48.4 | 0.0 |
| 09/2017 | 19345 | 71.0 | 28.7 | 0.3 | 624 | 61.4 | 38.1 | 0.5 | 2187021 | 62.4 | 37.6 | 0.0 |
| 10/2017 | 19123 | 71.4 | 28.3 | 0.3 | 634 | 61.7 | 37.9 | 0.5 | 2010559 | 67.3 | 32.7 | 0.0 |
| 11/2017 | 18746 | 69.1 | 30.6 | 0.3 | 607 | 60.6 | 38.9 | 0.5 | 3750020 | 71.4 | 28.6 | 0.0 |
| 12/2017 | 17331 | 67.6 | 32.1 | 0.3 | 606 | 63.0 | 36.5 | 0.5 | 2537571 | 50.1 | 49.9 | 0.0 |
| 01/2018 | 19014 | 69.7 | 30.0 | 0.3 | 629 | 61.7 | 37.7 | 0.6 | 2703535 | 59.2 | 40.8 | 0.0 |
| 02/2018 | 16806 | 68.2 | 31.5 | 0.3 | 618 | 61.7 | 37.4 | 1.0 | 2725997 | 54.2 | 45.8 | 0.0 |
| 03/2018 | 18433 | 68.0 | 31.6 | 0.3 | 617 | 60.3 | 39.1 | 0.6 | 2469085 | 56.7 | 43.3 | 0.0 |
| 04/2018 | 17771 | 67.5 | 32.3 | 0.2 | 598 | 59.5 | 39.6 | 0.8 | 3009583 | 58.9 | 41.1 | 0.0 |
| 05/2018 | 16739 | 67.6 | 32.0 | 0.4 | 602 | 58.5 | 40.7 | 0.8 | 2281952 | 47.6 | 52.4 | 0.0 |
| 06/2018 | 15178 | 67.4 | 32.4 | 0.2 | 601 | 58.2 | 41.3 | 0.5 | 2590198 | 71.4 | 28.6 | 0.0 |
| 07/2018 | 16225 | 70.6 | 29.2 | 0.1 | 624 | 59.9 | 39.4 | 0.6 | 1403892 | 54.9 | 45.1 | 0.0 |
| 08/2018 | 16360 | 67.7 | 32.1 | 0.2 | 638 | 58.9 | 40.4 | 0.6 | 1519633 | 53.3 | 46.7 | 0.0 |
| 09/2018 | 15948 | 67.6 | 32.3 | 0.1 | 624 | 59.5 | 40.1 | 0.5 | 1395174 | 54.2 | 45.8 | 0.0 |
| 10/2018 | 18377 | 65.5 | 34.3 | 0.2 | 649 | 59.0 | 40.1 | 0.9 | 1670604 | 50.7 | 49.3 | 0.0 |
| 11/2018 | 17574 | 66.4 | 33.5 | 0.1 | 669 | 58.9 | 40.4 | 0.7 | 2090182 | 47.7 | 52.3 | 0.0 |
| 12/2018 | 17283 | 66.8 | 33.0 | 0.2 | 652 | 58.1 | 41.4 | 0.5 | 1860983 | 58.0 | 42.0 | 0.0 |
| 01/2019 | 20819 | 65.9 | 34.0 | 0.1 | 695 | 59.4 | 40.3 | 0.3 | 2193122 | 53.7 | 46.3 | 0.0 |
| 02/2019 | 22835 | 64.2 | 35.8 | 0.1 | 769 | 55.9 | 43.7 | 0.4 | 3002673 | 55.1 | 44.9 | 0.0 |
| 03/2019 | 27341 | 66.7 | 33.2 | 0.1 | 817 | 54.0 | 45.3 | 0.7 | 3131218 | 58.7 | 41.3 | 0.0 |
| 04/2019 | 25084 | 65.1 | 34.7 | 0.2 | 814 | 54.8 | 44.7 | 0.5 | 2813407 | 54.1 | 45.9 | 0.0 |
| 05/2019 | 24604 | 65.5 | 34.4 | 0.1 | 810 | 53.7 | 45.7 | 0.6 | 2267324 | 56.7 | 43.3 | 0.0 |
| 06/2019 | 21406 | 65.3 | 34.6 | 0.1 | 786 | 54.5 | 44.9 | 0.6 | 1975834 | 63.9 | 36.1 | 0.0 |
| 07/2019 | 18346 | 63.2 | 36.7 | 0.1 | 724 | 56.4 | 43.1 | 0.6 | 1401804 | 50.6 | 49.4 | 0.0 |
| 08/2019 | 18136 | 61.8 | 38.0 | 0.1 | 723 | 58.1 | 41.4 | 0.6 | 1189524 | 48.8 | 51.2 | 0.0 |
| 09/2019 | 18981 | 65.2 | 34.7 | 0.1 | 753 | 57.0 | 42.5 | 0.5 | 1781059 | 65.7 | 34.3 | 0.0 |
| 10/2019 | 18449 | 67.2 | 32.7 | 0.2 | 760 | 57.4 | 42.2 | 0.4 | 1822944 | 51.0 | 49.0 | 0.0 |
| 11/2019 | 18206 | 67.9 | 32.0 | 0.1 | 745 | 57.9 | 41.7 | 0.4 | 1536385 | 51.8 | 48.2 | 0.0 |
| 12/2019 | 18325 | 67.6 | 32.4 | 0.0 | 737 | 58.1 | 41.8 | 0.1 | 1448610 | 58.9 | 41.1 | 0.0 |
| 01/2020 | 20617 | 70.7 | 29.3 | 0.0 | 746 | 59.7 | 40.2 | 0.1 | 1964350 | 48.4 | 51.6 | 0.0 |
| 02/2020 | 18036 | 68.4 | 31.5 | 0.1 | 734 | 56.9 | 42.5 | 0.5 | 1791510 | 51.8 | 48.2 | 0.0 |
| 03/2020 | 22149 | 66.4 | 33.5 | 0.1 | 726 | 57.2 | 42.6 | 0.3 | 4692857 | 15.8 | 84.2 | 0.0 |
| 04/2020 | 26276 | 70.7 | 29.2 | 0.1 | 745 | 59.2 | 40.5 | 0.3 | 4244794 | 29.0 | 70.9 | 0.0 |
| 05/2020 | 28875 | 75.4 | 24.5 | 0.1 | 757 | 57.5 | 42.1 | 0.4 | 4088297 | 36.6 | 63.4 | 0.0 |
| 06/2020 | 21074 | 69.8 | 30.2 | 0.0 | 705 | 56.6 | 43.0 | 0.4 | 2940719 | 30.9 | 69.1 | 0.0 |
| 07/2020 | 22582 | 68.4 | 31.6 | 0.1 | 775 | 55.2 | 44.1 | 0.6 | 6445781 | 20.2 | 79.8 | 0.0 |
| 08/2020 | 22133 | 70.7 | 29.2 | 0.1 | 785 | 54.1 | 45.0 | 0.9 | 4988837 | 28.7 | 71.3 | 0.0 |
| 09/2020 | 21407 | 69.3 | 30.6 | 0.1 | 729 | 57.2 | 42.1 | 0.7 | 3196672 | 26.2 | 73.8 | 0.0 |
| 10/2020 | 21435 | 66.8 | 33.2 | 0.0 | 751 | 54.6 | 45.0 | 0.4 | 2317326 | 26.6 | 73.4 | 0.0 |
| 11/2020 | 19777 | 66.4 | 33.6 | 0.1 | 773 | 53.8 | 45.5 | 0.6 | 2536469 | 22.6 | 77.4 | 0.0 |

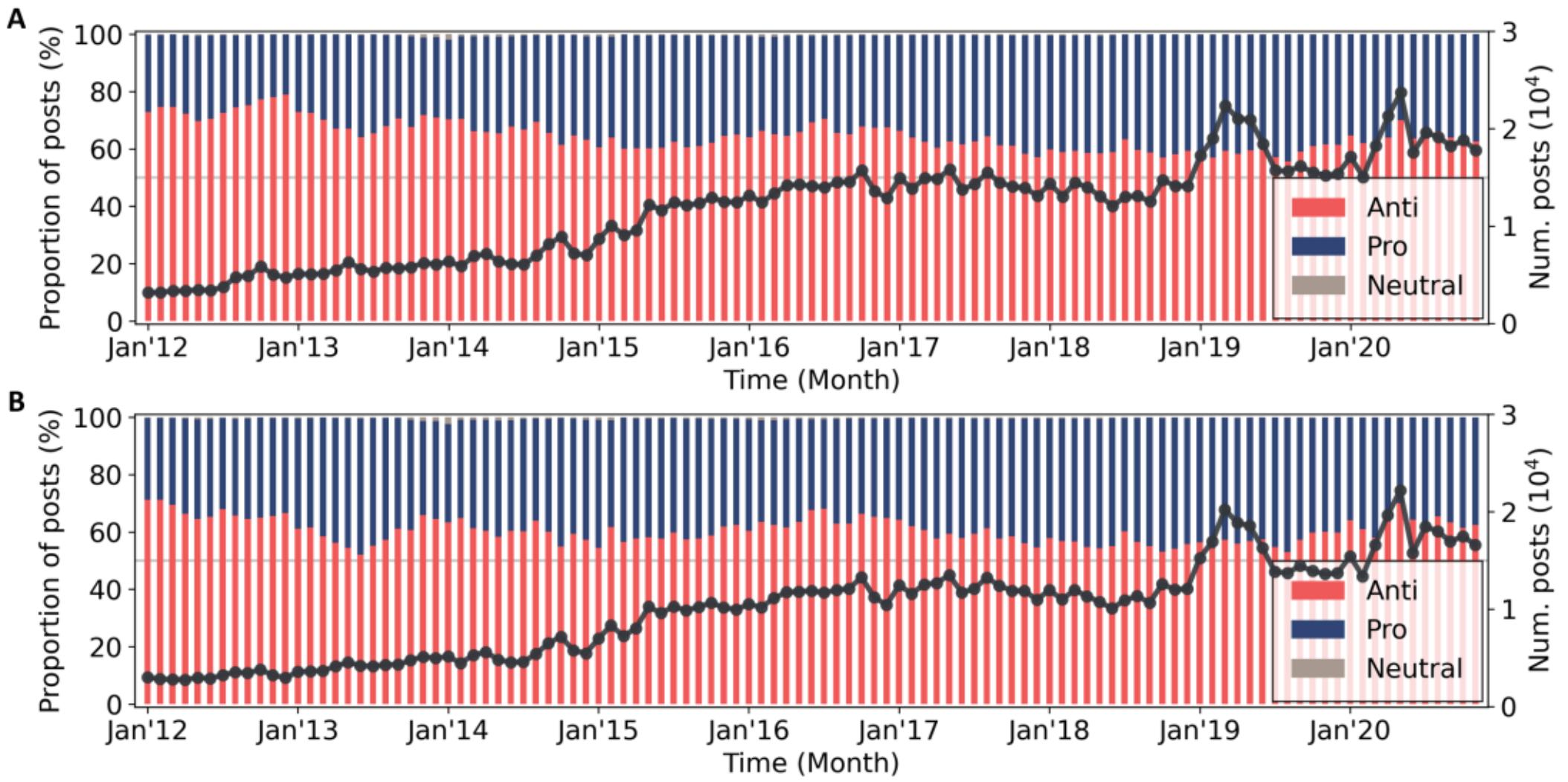

**Fig. S1**. **A**. The proportion of posts by group valence, excluding the top 10 vaccine groups in terms of the total number of posts generated. A stacked bar represents the proportions of Facebook posts created by pro-vaccine, anti-vaccine, and mixed/neutral groups in a given month (y-axis on the left). A dot indicates the total number of posts created by all vaccine groups in a given month (y-axis on the right). **B**. The proportion of posts by group valence, excluding the top 20 vaccine groups in terms of the total number of posts generated.

## S3.2 The Number of Active Groups

The 107-month average of the total number of active vaccine groups in a month was 485.9 (*SD* = 208.9, Median = 536). The average proportions of active anti-vaccine, pro-vaccine, and mixed/neutral groups were 58.4% (*SD* = 2.9%, Median = 58.3%), 40.9% (*SD* = 2.9%, Median = 40.7%), and 0.7% (*SD* = 0.3%, Median = 0.7%, *N* = 107), respectively. Table S2 shows the number of active groups in a month and the proportion of active anti-vaccine, pro-vaccine, and mixed/neutral groups.

## S3.3 The Number of Shares

The 107-month average of the total number of shares was 1933657.1 (*SD* = 1371819.3, Median = 1975834). The average proportions of shares received by anti- and pro-vaccine groups were 64.5% (*SD* = 17.0%, Median = 67.3%) and 35.5% (*SD* = 17.0%, Median = 32.7%), respectively. The average proportion of shares received by mixed/neutral groups was less than 0.01%. The total number of shares in each month and the proportion of shares received by anti-vaccine, pro-vaccine, and mixed/neutral groups are shown in Table S2.

## S3.4 The Number of Comments and Reactions

The 107-month average of the total number of comments in a month was 517223.3 (*SD* = 420892.7, Median = 509730). The average proportions of comments received by anti-vaccine, pro-vaccine, and mixed/neutral groups were 49.8% (*SD* = 15.0%, Median = 45.4%), 50.1% (*SD* = 15.0%, Median = 54.5%), and 0.1% (*SD* = 0.1%, Median = 0.0%, *N* = 107), respectively. The total number of comments received by all vaccine groups and the proportions of comments received by pro-vaccine, anti-vaccine, and mixed/neutral groups are visualized in Fig. S2A.

The 107-month average of the total number of reactions in a month was 4498344.6 (*SD* = 5160374.0, Median = 3630366). The average proportions of reactions received by anti-vaccine and pro-vaccine groups were 43.6% (*SD* = 14.7%, Median = 42.6%) and 56.4% (*SD* = 14.7%, Median = 57.4%), respectively. The average proportion of reactions received by mixed/neutral groups was less than 0.01%. The total number of reactions received by all vaccine groups and the proportions of comments received by pro-vaccine, anti-vaccine, and mixed/neutral groups are visualized in Fig. S2B.

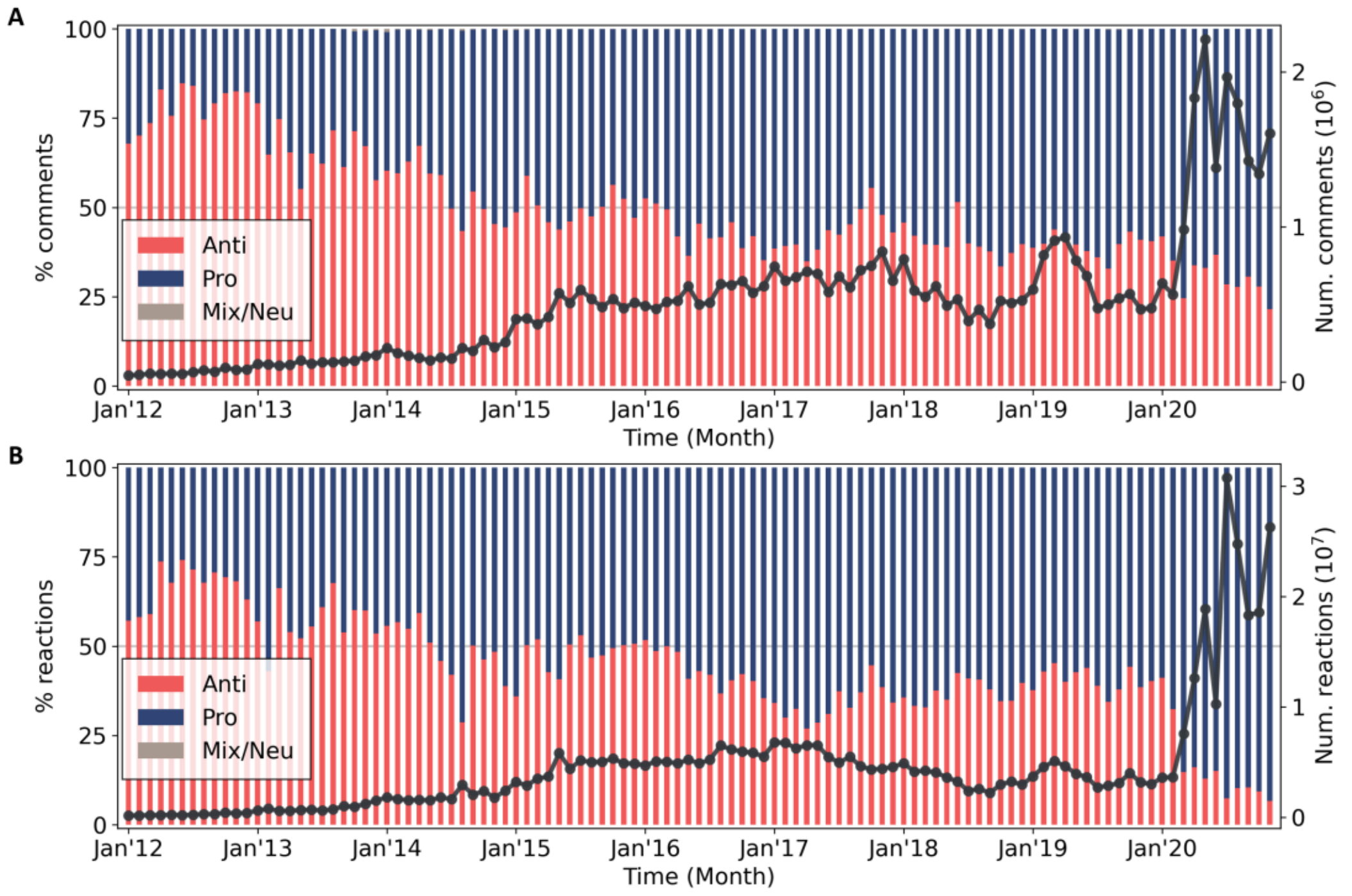

**Fig. S2**. **A.** Comments on vaccine content. A stacked bar represents the proportions of shares that pro-vaccine, anti-vaccine, mixed/neutral groups received in total in a given month (y-axis on the left). A dot indicates the total number of comments in a given month (y-axis on the right). **B.** Reaction to vaccine content.

## S3.5 Posts per Active Group in a Year

Table S3 shows the average number of posts in an active group in a year and the difference between active anti-vaccine and pro-vaccine groups. The results indicate that an anti-vaccine group created more posts than a pro-vaccine group in general. The 9-year average number of posts produced by anti-vaccine groups was also 2.4 times greater than pro-vaccine groups.

**Table S3**. The average number of posts per active vaccine group and the difference between pro- and anti-vaccine groups

| Year | Num. posts per active group (Pro-vaccine) | | | Num. posts per active group (Anti-vaccine) | | | Difference between pro and anti | |
|---|---|---|---|---|---|---|---|---|
| | Average (*SD*) | Median | *N* | Average (*SD*) | Median | *N* | MW test $U_1$ | KS test *D* |
| 2012 | 97.7 (134.4) | 34 | 125 | 404.0 (1316.8) | 38 | 177 | 9994.0 | 0.170 * |
| 2013 | 113.9 (221.3) | 26 | 181 | 374.4 (1199.6) | 38 | 219 | 17855.5 * | 0.125 |
| 2014 | 120.1 (247.7) | 25 | 229 | 368.7 (1048.4) | 53 | 252 | 24749.0 ** | 0.124 * |
| 2015 | 164.8 (383.5) | 27 | 311 | 369.7 (1530.8) | 48 | 410 | 55253.0 ** | 0.142 ** |
| 2016 | 160.6 (363.4) | 27 | 341 | 332.7 (1254.5) | 51 | 528 | 79317.0 ** | 0.114 ** |
| 2017 | 162.1 (431.0) | 21 | 404 | 292.5 (1221.1) | 42 | 585 | 104898.5 ** | 0.123 ** |
| 2018 | 144.2 (338.3) | 20 | 457 | 231.1 (918.2) | 30 | 603 | 127169.0 * | 0.070 |
| 2019 | 133.7 (361.9) | 13 | 650 | 238.3 (774.4) | 37 | 694 | 190222.0 *** | 0.154 *** |
| 2020 | 108.2 (283.2) | 13 | 685 | 237.8 (709.7) | 30 | 715 | 204730.0 *** | 0.162 *** |
| 9-year Average | 133.9 | | | 316.6 | | | | |

*Note.* Standard deviation in parenthesis. MW and KS stand for Mann-Whitney and Kolmogorov-Smirnov, respectively. $U_1$ is the MW statistics for the pro-vaccine sample of a given year. The MW statistics for the anti-vaccine sample of a given year, $U_2$, is calculated as $U_2 = N_1N_2 - U_1$. *$P < .05$, **$P < .01$, ***$P < .001$

## S3.6 Group Lifetime

The lifetime of a vaccine group was defined as the number of days between its first post and its last post. The lifetime can range between 0 (the first and last posts were posted on the same day) and 3,256 days (the first post on January 1st, 2012, and the last post on November 30th, 2020). A vaccine group with only one post was assigned a lifetime value of 0. The conversion from the number of days to the number of years was conducted by dividing the number of days by 365. Fig. 1E shows the distribution of lifetimes in the number of years.

The average lifetime of an anti-vaccine group was 1199.0 days ($SD$ = 1049.9, Median = 1004, $N$ = 1083) or 3.28 years. The average lifetime of a pro-vaccine group was 784.4 days ($SD$ = 974.8, Median = 333.5, $N$ = 1206) or 2.15 years. The difference between the anti- and pro-vaccine groups was significant (Mann-Whitney $U$ = 486188.5, $P < 0.0001$; Kolmogorov-Smirnov $D$ = 0.218, $P < 0.0001$).

# S4 Analysis: The Use of Information Sources

## S4.1 The Use of Information Sources

In total, 86.2% of all posts included at least one URL, while only 13.8% did not include any URL. When aggregated by vaccine groups, 82.7% (*SD* = 25.8%, *N* = 2328) of posts in a vaccine group included one or more URLs.

## S4.2 Shortened URLs

Link shortening services, such as bit.ly and tinyurl.com, receive URLs from their users and convert them to new shorter URLs that redirect to the original URLs. Facebook users often include shortened URLs in their social media content instead of original full URLs, and some of the information available in full URLs are removed during the process of link shortening. For example, bit.ly shortens https://en.wikipedia.org/wiki/Vaccine into https://bit.ly/3zqKgw1. In this example, while the original full URL includes information about the domain, i.e., wikipedia.org, its shortened version does not.

Hence, we identified shortened URLs, found their original full URLs, and extracted information about their actual domains. For this purpose, we built a list of 63 popular link shortening services by supplementing the list of 49 services reported by Yang et al. (*20*) and adding 14 more services to it (fb.me, is.gd, chng.it, tobtr.com, eepurl.com, cutt.ly, gf.me, hubs.ly, gates.ly, loom.ly, zurl.co, snip.ly, ed.gr, and m-gat.es) to cover all shortening services that are responsible for 100 or more URLs in our dataset. We detected shortened URLs based on this list, retrieved their original URLs by transmitting HTTP calls to the shortened URLs and capturing redirected full URLs, and then identified actual domains from the original URLs. Original URLs were not identifiable for some of the shortened URLs, and in these cases, domains of their link shortening services were considered as their actual domains.

**Table S4**. The domain list of link shortening services. 49 of the 63 services were adopted from Yang et al. (20)

| | | | | |
|---|---|---|---|---|
| bit.ly | j.mp | bitly.com | sc.mp | hubs.ly |
| dlvr.it | wapo.st | crfrm.us | gop.cm | gates.ly |
| liicr.nl | reut.rs | flip.it | crwd.fr | loom.ly |
| tinyurl.com | drudge.tw | mf.tt | zpr.io | zurl.co |
| goo.gl | shar.es | wp.me | scq.io | snip.ly |
| ift.tt | sumo.ly | voat.co | trib.in | ed.gr |
| ow.ly | rebrand.ly | zurl.co | owl.li | m-gat.es |
| fxn.ws | covfefe.bz | fw.to | fb.me | |
| buff.ly | trib.al | mol.im | is.gd | |
| back.ly | yhoo.it | read.bi | chng.it | |
| amzn.to | t.co | disq.us | tobtr.com | |
| nyti.ms | shr.lc | tmsnrt.rs | eepurl.com | |
| nyp.st | po.st | usat.ly | cutt.ly | |
| dailysign.al | dld.bz | aje.io | gf.me | |

## S4.3 Facebook Internal Sources and External Sources

We identified top-level domains of all URLs included in vaccine posts. For example, the top-level domain of an URL, https://www.cnn.com/health, is cnn.com. Each URL was classified into one of the two types of information sources based on its domain: Facebook internal sources and external sources.

Facebook internal sources were identified by detecting URLs whose top-level domains are the Facebook domains (e.g., facebook.com and fb.com). The inclusion of these Facebook domains in a post indicates that the post referenced or contained Facebook in-house content (such as photos, videos, and notes stored on Facebook), Facebook accounts (such as Facebook pages, groups, and individual profiles), or Facebook services (such as Facebook Watch) on a post. URLs that were not classified as Facebook internal sources were defined as external sources.

Fig. 2A shows the monthly average proportion of posts using at least one Facebook internal source in an active vaccine group and the monthly average proportion of posts using at least one external source in an active vaccine group.

Fig. S3 shows the average proportions of posts in an active pro- and anti-vaccine group in each year. First, Fig. S3A shows the average proportions of posts using at least one Facebook internal source in an active vaccine group each year. The proportion of posts using at least one Facebook internal source in an active pro-vaccine group had increased from 8.0% ($SD$ = 13.7%, Median = 2.3%, $N$ = 125) in 2012 to 55.7% ($SD$ = 35.6%, Median = 58.2%, $N$ = 685) in 2020. The difference between 2012 and 2020 was statistically significant (Mann-Whitney $U$ = 12703.0, $P < 0.0001$; Kolmogorov-Smirnov $D$ = 0.672, $P < 0.0001$). The proportion of posts using at least one Facebook internal source in an active anti-vaccine group had also increased from 9.5% (*SD* = 16.1%, Median = 3.9%, *N* = 177) in 2012 to 44.4% (*SD* = 31.2%, Median = 40.0%, *N* = 715) in 2020. The difference between 2012 and 2020 was statistically significant (Mann-Whitney *U* = 20684.5, $P < 0.0001$; Kolmogorov-Smirnov $D = 0.633$, $P < 0.0001$). The 9-year average proportion of posts using Facebook internal sources was 34.6% among pro-vaccine groups and 29.1% among anti-vaccine groups.

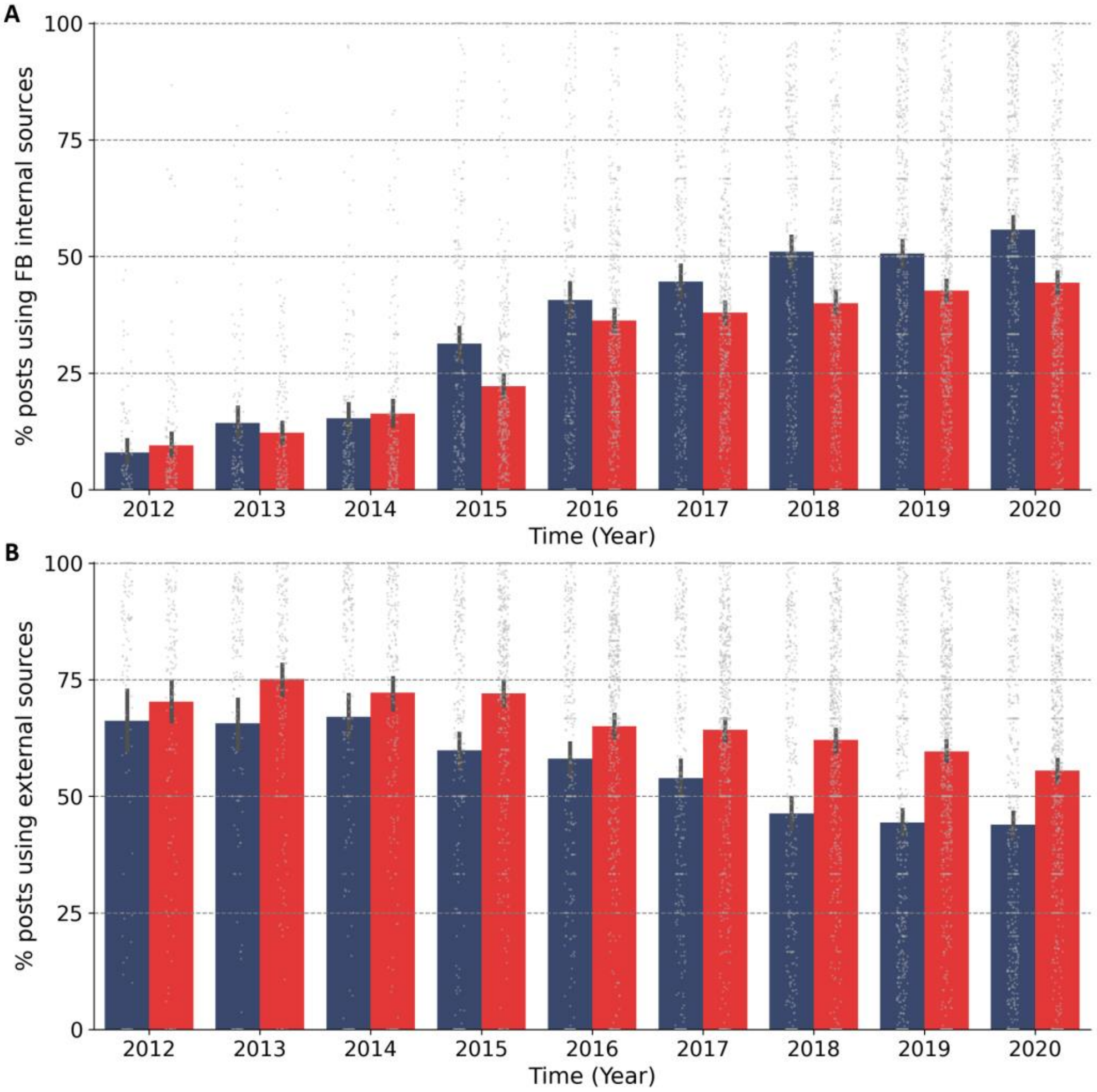

**Fig. S3 A.** The proportion of posts using at least one Facebook internal source in an active vaccine group in a year. Blue bars represent pro-vaccine groups, and red bars represent anti-vaccine groups. Error bars indicate 95% confidence intervals of the mean computed using bootstrapping. Each gray dot indicates an active vaccine group in a given year. **B**. The proportion of posts using at least one external source in an active vaccine group in a year.

Second, Fig. S3B shows the average proportions of posts using at least one external source in an active vaccine group each year. The proportion of posts using at least one external source in an active pro-vaccine group had decreased from 66.2% (*SD* = 36.2%, Median = 81.4%, *N* = 125) in 2012 to 43.9% (*SD* = 37.1%, Median = 42.0%, *N* = 685) in 2020. The difference between 2012 and 2020 was statistically significant (Mann-Whitney *U* = 29423.5, *P* < 0.0001; Kolmogorov-Smirnov *D* = 0.354, *P* < 0.0001). The proportion of posts using at least one external source in an active anti-vaccine group had also decreased from 70.3% (*SD* = 28.9%, Median = 78.2%, *N* = 177) in 2012 to 55.5% (*SD* = 31.0%, Median = 60.0%, *N* = 715) in 2020. The difference between 2012 and 2020 was statistically significant (Mann-Whitney *U* = 44645.5, *P* < 0.0001; Kolmogorov-Smirnov *D* = 0.303, *P* < 0.0001). The 9-year average proportion of posts using at least one external source was 56.1% among active pro-vaccine groups and 66.2% among active anti-vaccine groups.

### S4.3.1 Video Sources

Facebook and YouTube were the two most frequently used sources by vaccine groups, responsible for 35.2% of all URLs included in vaccine content. As an important case analysis examining the use of Facebook internal sources and external sources, we compared the proportions of video posts based on Facebook videos and YouTube videos. Facebook assigns one of the following labels to each post: photo, link, native_video, youtube, video, status, live_video_complete, live_video, live_video_scheduled, and album. We classified posts labeled native_video, video, live_video_complete, live_video, or live_video_scheduled into the "video posts using Facebook videos" category, and posts labeled youtube into the "video posts using YouTube videos" category. All posts in these two categories were defined as "video posts."

Videos were a major type of resources that vaccine groups utilized to support their narratives. Anti-vaccine groups tended to use videos more than pro-vaccine groups. First, the proportion of vaccine groups that created one or more video posts among all active groups each year is shown in Fig. S4A. The 9-year average proportion was 74.5% among anti-vaccine groups and 55.5% among pro-vaccine groups. Second, the proportion of video posts in an active vaccine group each year is displayed in Fig. S4B. The 9-year average was 14.5% among anti-vaccine groups and 7.3% among pro-vaccine groups.

The proportion of video posts using Facebook videos and YouTube videos in a vaccine group that created at least one video post each month is shown in Fig. 2B. Fig. S5 shows the proportion of video posts using Facebook videos and YouTube videos in a vaccine group that created at least one video post in a given year. Among pro-vaccine groups, the average proportion was 26.0% in 2012 (*SD* = 31.8%, *N* = 65) and 66.9% in 2020 (*SD* = 38.2%, *N* = 382). The difference between 2012 and 2020 was statistically significant (Mann-Whitney *U* = 5919.0, *P* < 0.0001; Kolmogorov-Smirnov *D* = 0.513, *P* < 0.0001). Among anti-vaccine groups, the average proportion was 16.2% in 2012 (*SD* = 24.8%, *N* = 123) and 63.3% in 2020 (*SD* = 32.1%, *N* = 537) on average. The difference was statistically significant (Mann-Whitney *U* = 9839.5, *P* < 0.0001; Kolmogorov-Smirnov *D* = 0.631, *P* < 0.0001).

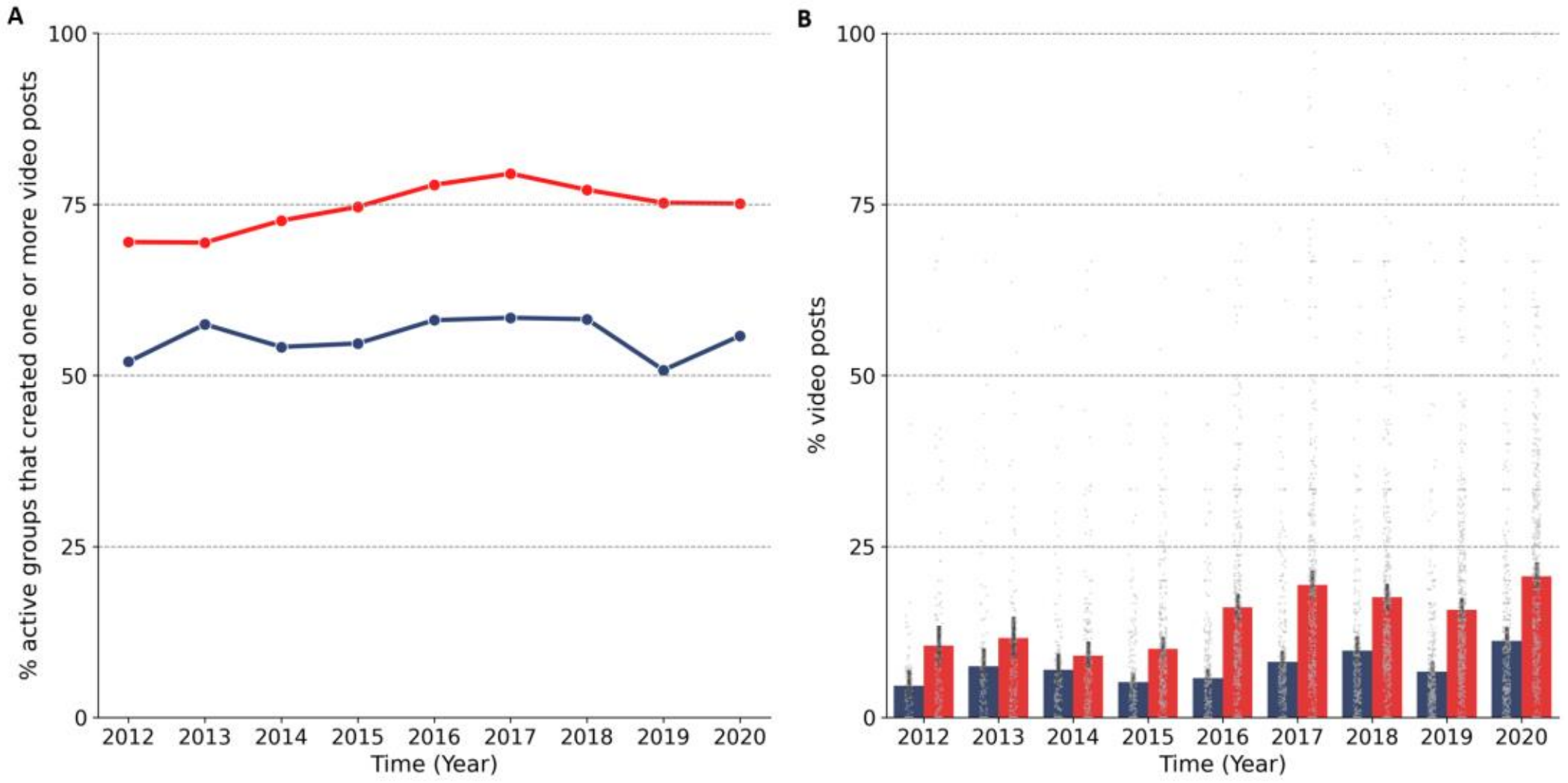

**Fig. S4. A**. The proportion of active groups that generated at least one video post among all active groups in a given year. Blue represents pro-vaccine groups, and red represents anti-vaccine groups. **B**. The proportion of video posts among all posts in a vaccine group in a given year. Error bars indicate 95% confidence intervals of the mean computed using bootstrapping. Each gray dot indicates an active vaccine group.

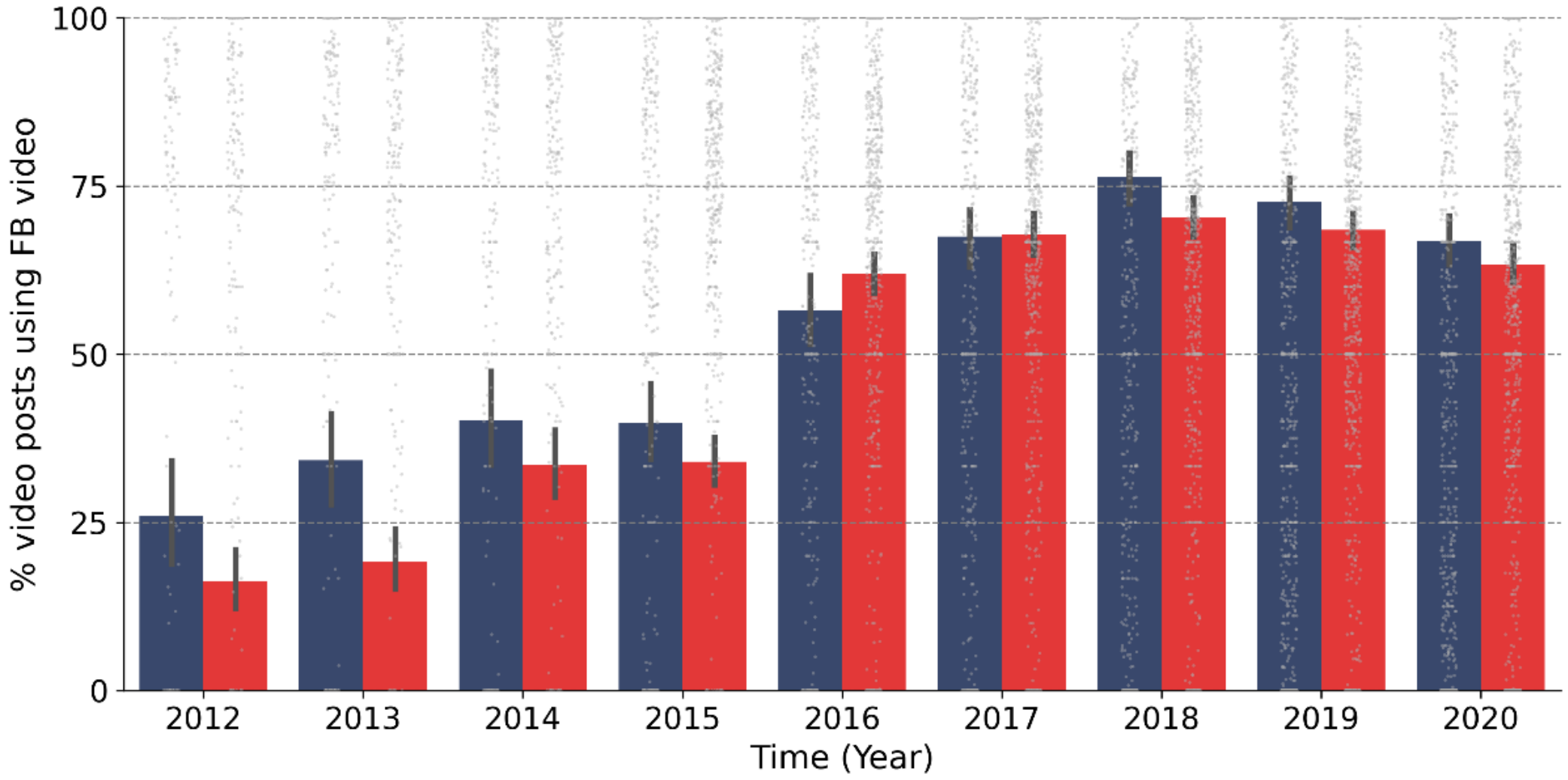

**Fig. S5**. The proportion of video posts using Facebook videos in a vaccine group that created at least one video post each year. Error bars indicate 95% confidence intervals of the mean computed using bootstrapping. Each gray dot indicates a vaccine group.

## S4.4 Subtypes of External Sources

All non-Facebook domains were defined as "external domains." URLs that were directed to webpages or resources within external domains were considered as indicating the use of external sources. We focused on the use of external sources that are known to supply information with low credibility (i.e., low credibility sources) and compared it with the use of other subtypes of external sources: government sources, social media sources, and news sources.

### S4.4.1 Government Sources

In total 15,707 domains were defined as government domains. These domains are included in at least one of the following lists: the .gov domains listed by gsa.gov (*21*), the U.S. government-managed non-.gov listed by search.gov (*22*), and the 87 international government domains (*23*). If the top-level domain of an URL was a government domain (for example, https://www.cdc.gov/vaccines/index.html), the URL was considered as indicating the use of a government source.

Fig. 2C shows the monthly average proportions of posts using one or more government sources in an active anti- or pro-vaccine group. The 107-month average was 3.6% (*SD* = 1.5%) for anti-vaccine groups and 4.5% (*SD* = 1.3%) for pro-vaccine groups.

As reported in Table S5, Mann-Whitney tests show that the proportion of posts using government sources was significantly greater in an active anti-vaccine group than an active pro-vaccine group between 2015 and 2020. The average proportions of posts using government sources in an active group each year are shown in Fig S6A.

**Table S5**. The proportion of posts using at least one government source in an active pro- and anti-vaccine groups

| Year | Pro-vaccine group | | | Anti-vaccine group | | | Difference between pro and anti | |
|---|---|---|---|---|---|---|---|---|
| | Avg. (%) (*SD*) | Median (%) | $N_1$ | Avg. (%) (*SD*) | Median (%) | $N_2$ | MW test $U_1$ | KS test *D* |
| 2012 | 6.4 (15.2) | 0.5 | 125 | 3.3 (9.5) | 0.2 | 177 | 10328.0 | 0.109 |
| 2013 | 5.0 (12.0) | 0.0 | 181 | 2.5 (5.6) | 0.0 | 219 | 19527.5 | 0.099 |
| 2014 | 3.6 (8.6) | 0.0 | 229 | 2.5 (7.5) | 0.0 | 252 | 28095.0 | 0.089 |
| 2015 | 4.2 (11.9) | 0.0 | 311 | 4.3 (10.3) | 1.0 | 410 | 55727.0 ** | 0.134 ** |
| 2016 | 4.2 (12.5) | 0.0 | 341 | 4.2 (10.3) | 0.7 | 528 | 77482.0 *** | 0.164 *** |
| 2017 | 2.9 (9.2) | 0.0 | 404 | 4.4 (10.3) | 0.8 | 585 | 93942.0 *** | 0.215 *** |
| 2018 | 2.8 (9.5) | 0.0 | 457 | 4.8 (10.3) | 0.5 | 603 | 109190.5 *** | 0.215 *** |
| 2019 | 3.3 (12.1) | 0.0 | 650 | 5.9 (11.7) | 1.8 | 694 | 156681.5 *** | 0.292 *** |
| 2020 | 3.3 (11.9) | 0.0 | 685 | 4.6 (11.1) | 0.6 | 715 | 185850.0 *** | 0.254 *** |

*Note*. Standard deviation in parenthesis. MW and KS stand for Mann-Whitney and Kolmogorov-Smirnov, respectively. $U_1$ is the MW statistics for the pro-vaccine sample of a given year. The MW statistics for the anti-vaccine sample of a given year, $U_2$, is calculated as $U_2 = N_1N_2 - U_1$. *$P < .05$, **$P < .01$, ***$P < .001$

Among pro-vaccine groups, the proportion of posts using government sources in an active pro-vaccine group decreased from 6.4% in 2012 (*SD* = 15.2%, *N* = 125) to 3.3% in 2020 (*SD* = 11.9%, *N* = 685), and the difference was statistically significant (Mann-Whitney *U* = 32551.0, *P* < 0.0001; Kolmogorov-Smirnov *D* = 0.230, *P* < 0.0001). Among anti-vaccine groups, the proportion of posts using government sources in an active anti-vaccine group increased from 2.5% (*SD* = 5.6%, *N* = 219) in 2013 to 4.6% (*SD* = 11.1%, *N* = 715) in 2020, and the difference was significant (*U* = 70099.0, *P* = 0.0064; *D* = 0.141, *P* = 0.0021).

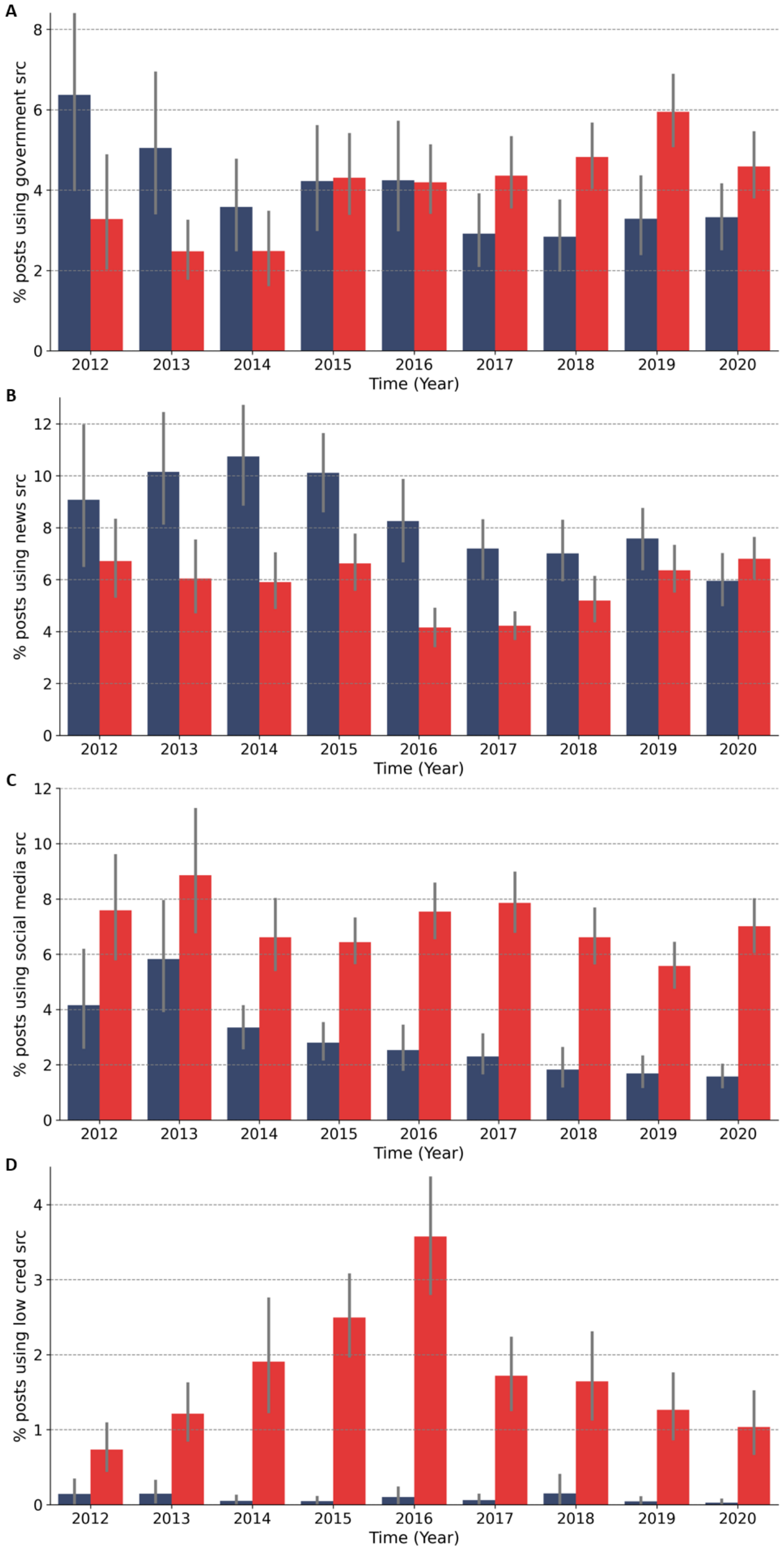

**Fig S6**. **A.** The proportion of posts using government sources in a group. A bar represents the proportions averaged among all active pro- or anti-vaccine groups in a given year. Error bars

indicate 95% confidence intervals of the mean computed using bootstrapping. Red represents anti-vaccine, and blue represents pro-vaccine. **B**. News sources. **C.** Social media sources. **D.** Low credibility sources.

### S4.4.2 News Sources

The news domain list includes "Hard news domains" in Bakshy et al. (*24*), "News media sites" in Yang et al. (*25*), "Newspapers" and "Digital-native news outlets" labeled by Pew Research Center (*26*), and "Green" and "Yellow" domains in Grinberg et al. (*27*). (Grinberg et al. categorized websites into six levels: green, yellow, orange, red, satire, and not applicable. According to their categorization criteria, the first five levels correspond to "reasonable and accountable journalism," "low quality journalism," "negligent or deceptive," "little regard for the truth," "self-described as satirical and affirmed as such by the annotators," respectively (*27*).) Domains that were also included in the low credibility domain list were excluded from the news domain list. If the top-level domain of an URL exists in the news domain list (e.g., https://www.cnn.com/health), the URL was considered as indicating the use of a news source.

Fig. 2D shows the monthly average proportions of posts using one or more news sources in an active anti- or pro-vaccine group. The 107-month average was 6.5% (*SD* = 1.6%) for anti-vaccine groups and 10.5% (*SD* = 2.6%) for pro-vaccine groups.

As reported in Table S6 and visualized in Fig S6B and also in Fig. 2D, the proportion of posts using news sources was significantly greater in a pro-vaccine group than in an anti-vaccine group between 2013 and 2015. In 2020, however, anti-vaccine groups (*M* = 6.8%, *SD* = 10.6%, Median = 2.8%, *N* = 715) surpassed pro-vaccine groups (*M* = 6.0%, *SD* = 12.7%, Median = 0.0%, *N* = 685) in terms of the proportion of posts using news sources, and the gap between anti- and pro-vaccine groups was statistically significant as shown in Table S6.

**Table S6**. The proportion of posts using at least one news source in an active pro- and anti-vaccine groups

| Year | Pro-vaccine group | | | Anti-vaccine group | | | Difference between pro and anti | |
|---|---|---|---|---|---|---|---|---|
| | Avg. (%) (*SD*) | Median (%) | $N_1$ | Avg. (%) (*SD*) | Median (%) | $N_2$ | MW test $U_1$ | KS test *D* |
| 2012 | 9.1 (15.6) | 2.8 | 125 | 6.7 (10.1) | 3.3 | 177 | 10870.5 | 0.100 |
| 2013 | 10.1 (15.6) | 4.0 | 181 | 6.0 (10.5) | 2.4 | 219 | 17545.5 * | 0.159 * |
| 2014 | 10.7 (15.0) | 4.6 | 229 | 5.9 (8.5) | 2.4 | 252 | 25169.5 ** | 0.197 *** |
| 2015 | 10.1 (12.7) | 5.2 | 311 | 6.6 (10.7) | 3.6 | 410 | 58035.5 * | 0.192 *** |
| 2016 | 8.3 (15.2) | 1.0 | 341 | 4.2 (8.8) | 1.3 | 528 | 82704.0 * | 0.187 *** |
| 2017 | 7.2 (11.1) | 0.8 | 404 | 4.2 (6.6) | 1.8 | 585 | 113527.0 | 0.154 *** |
| 2018 | 7.0 (12.2) | 0.0 | 457 | 5.2 (10.7) | 1.3 | 603 | 137617.0 | 0.090 * |
| 2019 | 7.6 (14.6) | 0.0 | 650 | 6.4 (11.7) | 2.6 | 694 | 210246.0 * | 0.130 *** |
| 2020 | 6.0 (12.7) | 0.0 | 685 | 6.8 (10.6) | 2.8 | 715 | 201180.5 *** | 0.203 *** |

*Note.* Standard deviation in parenthesis. MW and KS stand for Mann-Whitney and Kolmogorov-Smirnov, respectively. $U_1$ is the MW statistics for the pro-vaccine sample of a given year. The MW statistics for the anti-vaccine sample of a given year, $U_2$, is calculated as $U_2 = N_1N_2 - U_1$. *$P < .05$, **$P < .01$, ***$P < .001$

Among pro-vaccine groups, the proportion of posts using news sources in an active group significantly decreased from 10.7% in 2014 (*SD* = 15.0%, Median = 4.6%, *N* = 229) to 6.0% in 2020 (*SD* = 12.7%, Median = 0.0%, *N* = 685; Mann-Whitney *U* = 59389.5, *P* < 0.0001; Kolmogorov-Smirnov *D* = 0.241, *P* < 0.0001). For anti-vaccine groups, the proportion of posts using news sources displayed a significant increase between 2016 and 2020 from 4.2% (*SD* = 8.8%, Median = 1.3%, *N* = 528) in 2016 to 6.8% (*SD* = 10.6%, Median = 2.8%, *N* = 715) in 2020 (*U* = 165006.0, *P* < 0.0001; *D* = 0.158, *P* < 0.0001). These longitudinal changes are also visible in Fig S6B and Fig. 2D

### S4.4.3 Social Media Sources

The social media domain list includes the domains of social media sites that Pew Research Center identified as social media in one or more of the following reports they had published between 2012 and 2020 (*28–36*). The social media domain list consists of the domains of Twitter, YouTube, Instagram, TikTok, Reddit, Tumblr, Google+, Twitch, Vine, WhatsApp, and Snapchat. If the top-level domain of an URL exists in this list, the URL was considered as indicating the use of a social media source.

Fig. 2E shows the monthly average proportions of posts using one or more social media sources in an active anti- or pro-vaccine group. The 107-month average was 7.5% (*SD* = 1.4%) for anti-vaccine groups and 3.2% (*SD* = 1.6%) for pro-vaccine groups.

As reported in Table S7 and displayed in Fig S6C and Fig 2E, the proportion of posts using social media sources was significantly greater in an anti-vaccine group than a pro-vaccine group in every year between 2012 and 2020. Among pro-vaccine groups, the proportion of posts using social media sources in an active group was 4.2% (*SD* = 10.8%, Median = 0.0%, *N* = 125) in 2012 and 5.8% (*SD* = 13.6%, Median = 0.9%, *N* = 181) in 2013 on average, and it significantly decreased between 2013 and 2020 to 1.6% (*SD* = 5.4%, Median = 0.0%, *N* = 685; Mann-Whitney *U* = 44074.5, *P* < 0.0001; Kolmogorov-Smirnov *D* = 0.272, *P* < 0.0001). Among anti-vaccine groups, the average proportion of posts using social media sources in a group fluctuated between 5.6% (in 2019, *SD* = 10.8%, *N* = 694) and 8.9% (in 2013, *SD* = 16.8%, *N* = 219).

**Table S7**. The proportion of posts using at least one social media source in an active pro- and anti-vaccine groups

| Year | Pro-vaccine group | | | Anti-vaccine group | | | Difference between pro and anti | |
|---|---|---|---|---|---|---|---|---|
| | Avg. (%) (*SD*) | Median (%) | $N_1$ | Avg. (%) (*SD*) | Median (%) | $N_2$ | MW test $U_1$ | KS test *D* |
| 2012 | 4.2 (10.8) | 0.0 | 125 | 7.6 (13.3) | 2.9 | 177 | 8541.0 *** | 0.191 ** |
| 2013 | 5.8 (13.6) | 0.9 | 181 | 8.9 (16.8) | 3.2 | 219 | 16184.0 *** | 0.168 ** |
| 2014 | 3.4 (5.9) | 0.0 | 229 | 6.6 (10.8) | 2.4 | 252 | 22371.5 *** | 0.187 *** |
| 2015 | 2.8 (5.7) | 0.0 | 311 | 6.4 (8.4) | 3.9 | 410 | 44580.0 *** | 0.295 *** |
| 2016 | 2.5 (7.4) | 0.0 | 341 | 7.5 (11.4) | 4.0 | 528 | 56917.0 *** | 0.337 *** |
| 2017 | 2.3 (7.3) | 0.0 | 404 | 7.6 (12.9) | 3.4 | 585 | 76990.0 *** | 0.352 *** |
| 2018 | 1.8 (7.5) | 0.0 | 457 | 6.6 (12.7) | 2.2 | 603 | 88730.5 *** | 0.316 *** |
| 2019 | 1.7 (7.1) | 0.0 | 650 | 5.7 (10.8) | 1.9 | 694 | 145920.0 *** | 0.340 *** |
| 2020 | 1.6 (5.4) | 0.0 | 685 | 7.0 (12.5) | 2.1 | 715 | 155185.5 *** | 0.344 *** |

*Note*. Standard deviation in parenthesis. MW and KS stand for Mann-Whitney and Kolmogorov-Smirnov, respectively. $U_1$ is the MW statistics for the pro-vaccine sample of a given year. The MW statistics for the anti-vaccine sample of a given year, $U_2$, is calculated as $U_2 = N_1N_2 - U_1$. **P* < .05, ***P* < .01, ****P* < .001

### S4.4.4 Low Credibility Sources

The low credibility domain list includes (a) the domains categorized as "Black," "Red," or "Orange" sources by Grinberg et al. (*27*), and (b) the domains identified as "very low credibility" and "low credibility" by Media Bias and Fact Check, an “independent website that rates the bias, factual accuracy, and credibility of media sources” (*37*). According to Grinberg et al., the Black sources were taken from previous lists of domains reported by Guess et al. (*38*), Allcott and Gentzkow (*39*), Snopes.com, and Buzzfeed; and the Red and Orange sources are websites that they evaluated as “negligent or deceptive” and “little regard for the truth” (*27*). If the top-level domain of an URL was included in this list, the URL was considered as indicating the use of a low credibility source.

Fig. 2F shows the monthly average proportions of posts using one or more low credibility sources in an active anti- or pro-vaccine group. The 107-month average was 1.8% ($SD$ = 1.1%) for anti-vaccine groups and 0.1% ($SD$ = 0.1%) for pro-vaccine groups.

As reported in Table S8 and depicted in Fig S6D and Fig 2F, the proportion of posts using low credibility sources was significantly greater in an anti-vaccine group than a pro-vaccine group every year between 2012 and 2020. Among anti-vaccine groups, the proportion of posts using low credibility sources in a group increased from 0.7% ($SD$ = 2.1%, $N$ = 177) in 2012 to 3.6% ($SD$ = 9.0%, $N$ = 528) in 2016 (Mann-Whitney $U$ = 35034.0, $P < 0.0001$; Kolmogorov-Smirnov $D$ = 0.264, $P < 0.0001$) and then significantly decreased between 2016 and 2020 to 1.0% ($SD$ = 5.8%, $N$ = 715; $U$ = 138178.0, $P < 0.0001$; $D$ = 0.246, $P < 0.0001$).

**Table S8**. The proportion of posts using at least one low credibility source in an active pro- and anti-vaccine groups

| Year | Pro-vaccine group | | | Anti-vaccine group | | | Difference between pro and anti | |
|---|---|---|---|---|---|---|---|---|
| | Avg. (%) (*SD*) | Median (%) | $N_1$ | Avg. (%) (*SD*) | Median (%) | $N_2$ | MW test $U_1$ | KS test $D$ |
| 2012 | 0.1 (0.9) | 0.0 | 125 | 0.7 (2.1) | 0.0 | 177 | 8076.0 *** | 0.269 *** |
| 2013 | 0.1 (0.9) | 0.0 | 181 | 1.2 (2.8) | 0.0 | 219 | 12730.0 *** | 0.355 *** |
| 2014 | 0.1 (0.4) | 0.0 | 229 | 1.9 (6.2) | 0.0 | 252 | 16903.5 *** | 0.411 *** |
| 2015 | 0.0 (0.3) | 0.0 | 311 | 2.5 (6.0) | 0.5 | 410 | 32518.0 *** | 0.482 *** |
| 2016 | 0.1 (0.9) | 0.0 | 341 | 3.6 (9.0) | 0.2 | 528 | 48859.0 *** | 0.446 *** |
| 2017 | 0.1 (0.6) | 0.0 | 404 | 1.7 (5.9) | 0.0 | 585 | 74576.0 *** | 0.362 *** |
| 2018 | 0.2 (2.4) | 0.0 | 457 | 1.6 (7.4) | 0.0 | 603 | 92424.0 *** | 0.323 *** |
| 2019 | 0.0 (0.6) | 0.0 | 650 | 1.3 (5.9) | 0.0 | 694 | 158677.0 *** | 0.296 *** |
| 2020 | 0.0 | 0.0 | 685 | 1.0 | 0.0 | 715 | 177020.5 *** | 0.276 *** |

| | |
|---|---|
| (0.4) | (5.9) |

*Note.* Standard deviation in parenthesis. MW and KS stand for Mann-Whitney and Kolmogorov-Smirnov, respectively. $U_1$ is the MW statistics for the pro-vaccine sample of a given year. The MW statistics for the anti-vaccine sample of a given year, $U_2$, is calculated as $U_2 = N_1N_2 - U_1$. *$P$ < .05, **$P$ < .01, ***$P$ < .001

## S5 Analysis: User Engagement and Information Sources

We examined the association between the use of information sources in a post and the level of user engagement with the post in pro-vaccine and anti-vaccine groups. Three different variables represented the level of user engagement with a post: the number of shares of a post (*M* = 123.3, *SD* = 2139.8, *N* = 1,678,351), the number of comments to a post (*M* = 33.0, *SD* = 330.6, *N* = 1,678,351), and the number of reactions to a post (*M* = 286.8, *SD* = 6196.9, *N* = 1,678,351). The number of reactions was computed by summing eight different reaction counters of a post: like, love, wow, haha, sad, angry, thankful, and care. SciPy package (version 1.6.1) for the Python programming language was used for all analyses.

For the posts using at least one information source, which corresponds to 86.2% (*N* = 1,446,275) of all posts in our data, we fit a negative binomial regression model predicting each of the three user engagement variables. The sample means of the three dependent variables were as follows: the number of shares of a post (*M* = 140.9, *SD* = 2284.9, *N* = 1,446,275), the number of comments to a post (*M* = 34.7, *SD* = 349.9, *N* = 1,446,275), and the number of reactions to a post (*M* = 320.3, *SD* = 6344.6, *N* = 1,446,275). 71.4% of the posts were shared at least once. 63.4% of the posts received at least one comment, and 91.0% received at least one reaction.

The regression models included six covariates: (1) Vaccine opposition (a binary variable indicating whether the vaccine group that created the post is an anti-vaccine group (2) subscriber (a counting variable indicating the number of Facebook users subscribed to the vaccine group), (3) posting time (an integer variable ranging from 0 (January 2012) to 106 (November 2020)), (4) video (a binary variable indicating whether a post includes a video), (5) photo (a binary variable indicating whether a post includes a photo), (6) Facebook group (a binary variable indicating whether the vaccine group is a Facebook group or a Facebook page). The second covariate, the number of subscribers, was measured at the time of data collection. (CT also reports the number of subscribers measured at the time of post creation, but this metric was not used for the current analysis because, according to the company's announcement (*40*), it does not provide accurate information for posts published before August 2017.)

All standard errors were clustered at the vaccine group level, and all models were estimated with cluster-robust standard error at the vaccine group level. We computed incidence rate ratios (*IRR*s) by exponentiating the coefficients of each negative binomial regression model. Regression analysis results for shares, comments, and reactions are reported in Fig 3B, Table S9, and Table S10, respectively.

**Table S9.** Results of a negative binomial model estimating the number of comments to a post

| | coef | IRR | 95% CI | std err | z | P > \|z\| |
|---|---|---|---|---|---|---|
| FB internal source | 0.5524 | 1.7374 | [1.2145, 2.4856] | 0.183 | 3.024 | 0.002 |
| Government source | -0.5585 | 0.5721 | [0.4604, 0.7108] | 0.111 | -5.041 | <0.001 |
| News source | 0.1169 | 1.1240 | [0.8648, 1.4609] | 0.134 | 0.874 | 0.382 |
| Social media source | -1.1875 | 0.3050 | [0.2048, 0.4543] | 0.203 | -5.841 | <0.001 |
| Low credibility source | 0.5303 | 1.6995 | [1.0963, 2.6345] | 0.224 | 2.371 | 0.018 |
| Other source | -0.4642 | 0.6286 | [0.5064, 0.7804] | 0.110 | -4.208 | <0.001 |
| Vaccine opposition | -0.5235 | 0.5924 | [0.4040, 0.8688] | 0.195 | -2.680 | 0.007 |
| Photo | -0.2140 | 0.8074 | [0.5922, 1.1007] | 0.158 | -1.353 | 0.176 |
| Video | 0.3534 | 1.4238 | [1.1118, 1.8234] | 0.126 | 2.800 | 0.005 |
| Subscriber | 1.254e-06 | 1.0000 | [1.0000, 1.0000] | 4.64e-07 | 2.702 | 0.007 |
| Month | 0.0089 | 1.0089 | [1.0049, 1.0130] | 0.002 | 4.375 | <0.001 |
| FB group (ref: FB page) | -1.1075 | 0.3304 | [0.2186, 0.4994] | 0.211 | -5.254 | <0.001 |
| Intercept | 2.6365 | 13.9644 | [7.1223,27.3793] | 0.344 | 7.675 | <0.001 |

*Note.* Results of a negative binomial model estimating the number of comments of a post as a function of variables shown in the leftmost column. Coef, regression coefficient; IRR, incidence rate ratio; 95% CI, 95% confidence interval of IRR; z, z score. All standard errors were clustered at the vaccine group level, and all models were estimated with cluster-robust standard error at the vaccine group level. $N$ = 1,446,275

**Table S10.** Results of a negative binomial model estimating the number of reactions to a post

| | coef | IRR | 95% CI | std err | z | P > \|z\| |
|---|---|---|---|---|---|---|
| FB internal source | 0.1814 | 1.1989 | [0.8647, 1.6623] | 0.167 | 1.088 | 0.277 |
| Government source | -0.8321 | 0.4351 | [0.3259, 0.5810] | 0.147 | -5.643 | <0.001 |
| News source | 0.0425 | 1.0435 | [0.8271, 1.3164] | 0.119 | 0.359 | 0.720 |
| Social media source | -0.9640 | 0.3814 | [0.2571, 0.5657] | 0.201 | -4.792 | <0.001 |
| Low credibility source | 0.5915 | 1.8066 | [1.2764, 2.5571] | 0.177 | 3.337 | 0.001 |
| Other source | -0.2992 | 0.7414 | [0.5988, 0.9179] | 0.109 | -2.746 | 0.006 |
| Vaccine opposition | -0.2775 | 0.7577 | [0.5292, 1.0849] | 0.183 | -1.515 | 0.130 |
| Photo | 0.5376 | 1.7120 | [1.3152, 2.2284] | 0.135 | 3.997 | <0.001 |
| Video | 0.3167 | 1.3726 | [1.0633, 1.7719] | 0.130 | 2.431 | 0.015 |
| Subscriber | 1.796e-06 | 1.0000 | [1.0000, 1.0000] | 5.23e-07 | 3.437 | 0.001 |
| Month | 0.0063 | 1.0063 | [1.0019, 1.0107] | 0.002 | 2.812 | 0.005 |
| FB group (ref: FB page) | -2.0302 | 0.1313 | [0.0811, 0.2127] | 0.246 | -8.251 | <0.001 |
| Intercept | 4.2118 | 67.4765 | [32.7745,138.9215] | 0.368 | 11.431 | <0.001 |

*Note*. Results of a negative binomial model estimating the number of reactions of a post as a function of variables shown in the leftmost column. Coef, regression coefficient; IRR, incidence rate ratio; 95% CI, 95% confidence interval of IRR; z, z score. All standard errors were clustered at the vaccine group level, and all models were estimated with cluster-robust standard error at the vaccine group level. *N* = 1,446,275

## S5.1 Robustness to Covariate Specification

We checked the robustness of the aforementioned findings by estimating the association between the use of information sources in a post and the levels of user engagement with it while controlling for different sets of covariates. As shown in Table S11, the associations were very robust to the inclusion of vaccine opposition and month in the regression models.

**Table S11.** Information sources and the number of shares, robustness to vaccine opposition and month

| | Share | | | Comment | | |
|---|---|---|---|---|---|---|
| | Model 1 | Model 2 | Model 3 | Model 1 | Model 2 | Model 3 |
| FB internal | 0.5155 | 0.4906 | 0.3669 | 0.5753 ** | 0.6525 ** | 0.5524 ** |
| Government | -0.6235 *** | -0.6468 *** | -0.7424 *** | -0.5153 *** | -0.4748 *** | -0.5585 *** |
| News | 0.0193 | 0.0375 | -0.0518 | 0.2136 | 0.1910 | 0.1169 |
| Social media | -0.7910 *** | -0.8038 *** | -0.7948 *** | -1.2878 *** | -1.2496 *** | -1.1875 *** |
| Low credibility | 1.0490 *** | 0.9204 *** | 0.8649 *** | 0.2951 | 0.5798 * | 0.5303 * |
| Other | -0.1728 | -0.2425 * | -0.2631 * | -0.5434 *** | -0.4436 *** | -0.4642 *** |
| Vaccine opposition | | 0.3416 | 0.4410 * | | -0.5780 ** | -0.5235 ** |
| Month | | | 0.0108 *** | | | 0.0089 *** |
| FB group (ref: page) | -2.8730 *** | -2.9509 *** | -3.1031 *** | -1.2768 *** | -1.1032 *** | -1.1075 *** |
| Subscriber | 1.83e-6 *** | 1.90e-6 *** | 1.99e-6 *** | 1.36e-6 ** | 1.21e-6 ** | 1.25e-6 ** |
| Photo | 0.2970 | 0.3711 | 0.4158 | -0.0740 | -0.1917 | -0.2140 |
| Video | 0.5414 ** | 0.5162 ** | 0.4960 ** | 0.3957 ** | 0.4445 ** | 0.3534 ** |

| | Reaction | | |
|---|---|---|---|
| | Model 1 | Model 2 | Model 3 |
| FB internal | 0.2219 | 0.2393 | 0.1814 |
| Government | -0.8115 *** | -0.7813 *** | -0.8321 *** |
| News | 0.1136 | 0.0948 | 0.0425 |
| Social media | -1.0190 *** | -0.9992 *** | -0.9640 *** |
| Low credibility | 0.4758 * | 0.6082 ** | 0.5915 ** |
| Other | -0.3402 ** | -0.2921 ** | -0.2992 ** |
| Vaccine opposition | | -0.3227 | -0.2775 |
| Month | | | 0.0063 ** |
| FB group (ref: page) | -2.0969 *** | -2.0175 *** | -2.0302 *** |
| Subscriber | 1.83e-6 *** | 1.74e-6 ** | 1.80e-6 ** |
| Photo | 0.5950 *** | 0.5349 *** | 0.5376 *** |
| Video | 0.3558 * | 0.3688 ** | 0.3167 * |

*Note*. Results of negative binomial models estimating the number of shares, comments, and reactions of a post as a function of variables shown in the leftmost column. The values are regression coefficients. All standard errors were clustered at the vaccine group level, and all models were estimated with cluster-robust standard error at the vaccine group level. *N* = 1,446,275. **P* < .05, ***P* < .01, ****P* < .001

## S5.2 Subgroup Analysis: Pro-vaccine and Anti-vaccine

To examine if the findings reported in S5 are consistent within a subset of pro-vaccine groups or anti-vaccine groups, we estimated the association between the use of information sources and user engagement among pro-vaccine and anti-vaccine groups separately. The regression models estimated for this robustness check were identical to those in S5, except that one variable, vaccine opposition was not included in the models since it does not vary within each subgroup. In total, 6 models (3 user engagement variables × 2 subgroups) were tested for this check. All standard errors were clustered at the vaccine group level, and all models were estimated with cluster-robust standard error at the vaccine group level.

Even when the models were estimated for pro-vaccine groups and anti-vaccine groups separately, as shown in Table S12, Table S13, and Table S14, the associations of the use of low credibility sources with the three user engagement variables were still positive and, in 4 out of 6 cases, still statistically significant. Regarding the use of social media sources in a post, all associations were still negative and significant even within the subsets of pro-vaccine groups and anti-vaccine groups. The associations between the use of government sources and user engagement variables were all still negative and, in 5 out of 6 cases, maintained their statistical significance.

**Table S12.** Results of negative binomial models estimating the number of shares of posts created by pro- and anti-vaccine groups

| | Posts by pro-vaccine groups | | Posts by anti-vaccine groups | |
|---|---|---|---|---|
| | IRR [95% CI] | P > \|z\| | IRR [95% CI] | P > \|z\| |
| FB internal | 1.2204 [0.4649, 3.2035] | 0.6858 | 1.5618 [0.9215, 2.6471] | 0.0976 |
| Government | 0.3058 [0.1367, 0.6842] | 0.0039 | 0.7470 [0.6407, 0.8709] | 0.0002 |
| News | 1.0156 [0.6153, 1.6764] | 0.9517 | 1.0816 [0.8557, 1.3670] | 0.5117 |
| Social media | 0.5749 [0.3570, 0.9257] | 0.0228 | 0.4568 [0.3245, 0.6431] | <0.0001 |
| Low credibility | 1.2084 [0.6687, 2.1838] | 0.5306 | 2.5538 [1.7515, 3.7236] | <0.0001 |
| Other | 0.5926 [0.4311, 0.8144] | 0.0013 | 0.9171 [0.7470, 1.1259] | 0.4081 |
| Photo | 2.2319 [0.7313, 6.8120] | 0.1580 | 1.2386 [0.7497, 2.0464] | 0.4019 |
| Video | 2.0633 [1.4187, 3.0007] | 0.0001 | 1.7585 [1.2226, 2.5291] | 0.0023 |
| Subscriber | 1.0000 [1.0000, 1.0000] | 0.1527 | 1.0000 [1.0000, 1.0000] | <0.0001 |
| Month | 1.0096 [1.0023, 1.0170] | 0.0102 | 1.0117 [1.0056, 1.0178] | 0.0001 |
| FB group (ref: FB page) | 0.0207 [0.0080, 0.0535] | <0.0001 | 0.0554 [0.0400, 0.0768] | <0.0001 |
| Intercept | 23.2327 [6.5397,82.5350] | <0.0001 | 16.3193 [8.3828,31.7700] | <0.0001 |
| *N* | 434,359 | | 1,011,916 | |

*Note*. Results of a negative binomial model estimating the number of shares of a post as a function of variables shown in the leftmost column. Coef, regression coefficient; IRR, incidence rate ratio; 95% CI, 95% confidence interval of IRR; All standard errors were clustered at the vaccine group level, and all models were estimated with cluster-robust standard error at the vaccine group level.

**Table S13.** Results of negative binomial models estimating the number of comments of posts created by pro- and anti-vaccine groups

| | Posts by pro-vaccine groups | | Posts by anti-vaccine groups | |
|---|---|---|---|---|
| | IRR [95% CI] | P > \|z\| | IRR [95% CI] | P > \|z\| |
| FB internal | 1.4607 [0.7164, 2.9784] | 0.2972 | 1.9072 [1.2732, 2.8567] | 0.0017 |
| Government | 0.4448 [0.1544, 1.2808] | 0.1333 | 0.8043 [0.6614, 0.9780] | 0.0290 |
| News | 1.0254 [0.7373, 1.4261] | 0.8816 | 1.2180 [0.9461, 1.5680] | 0.1259 |
| Social media | 0.4508 [0.3459, 0.5875] | <0.0001 | 0.2908 [0.1879, 0.4499] | <0.0001 |
| Low credibility | 1.6479 [1.0223, 2.6563] | 0.0403 | 1.6604 [1.0665, 2.5853] | 0.0248 |
| Other | 0.5878 [0.4450, 0.7765] | 0.0002 | 0.6537 [0.5081, 0.8411] | 0.0009 |
| Photo | 0.8717 [0.3771, 2.0147] | 0.7480 | 0.7601 [0.5440, 1.0621] | 0.1080 |
| Video | 1.3771 [1.0343, 1.8335] | 0.0284 | 1.5300 [1.1564, 2.0243] | 0.0029 |
| Subscriber | 1.0000 [1.0000, 1.0000] | 0.2631 | 1.0000 [1.0000, 1.0000] | 0.0018 |
| Month | 1.0098 [1.0035, 1.0162] | 0.0024 | 1.0089 [1.0040, 1.0138] | 0.0003 |
| FB group (ref: FB page) | 0.2815 [0.1189, 0.6668] | 0.0040 | 0.3876 [0.2603, 0.5773] | <0.0001 |
| Intercept | 17.4444 [6.1170,49.7482] | <0.0001 | 6.5279 [3.5485,12.0090] | <0.0001 |
| *N* | 434,359 | | 1,011,916 | |

*Note*. Results of a negative binomial model estimating the number of comments of a post as a function of variables shown in the leftmost column. Coef, regression coefficient; IRR, incidence rate ratio; 95% CI, 95% confidence interval of IRR; All standard errors were clustered at the vaccine group level, and all models were estimated with cluster-robust standard error at the vaccine group level.

**Table S14.** Results of negative binomial models estimating the number of reactions of posts created by pro- and anti-vaccine groups

| | Posts by pro-vaccine groups | | Posts by anti-vaccine groups | |
|---|---|---|---|---|
| | IRR [95% CI] | P > \|z\| | IRR [95% CI] | P > \|z\| |
| FB internal | 1.3063 [ 0.7652, 2.2301] | 0.3276 | 1.2183 [ 0.8424, 1.7619] | 0.2943 |
| Government | 0.2438 [ 0.1329, 0.4472] | <0.0001 | 0.6587 [ 0.5560, 0.7804] | <0.0001 |
| News | 1.1686 [ 0.7838, 1.7422] | 0.4446 | 1.0794 [ 0.8472, 1.3754] | 0.5363 |
| Social media | 0.6197 [ 0.4432, 0.8665] | 0.0051 | 0.3612 [ 0.2369, 0.5507] | <0.0001 |
| Low credibility | 1.1988 [ 0.7062, 2.0351] | 0.5018 | 1.6983 [ 1.1477, 2.5128] | 0.0081 |
| Other | 0.7198 [ 0.5569, 0.9302] | 0.0120 | 0.7565 [ 0.5910, 0.9684] | 0.0267 |
| Photo | 1.8150 [ 0.8804, 3.7416] | 0.1063 | 1.5281 [ 1.1473, 2.0351] | 0.0037 |
| Video | 1.4134 [ 1.0923, 1.8284] | 0.0084 | 1.4227 [ 1.0798, 1.8745] | 0.0122 |
| Subscriber | 1.0000 [ 1.0000, 1.0000] | 0.1463 | 1.0000 [ 1.0000, 1.0000] | 0.0002 |
| Month | 1.0062 [ 0.9988, 1.0136] | 0.0991 | 1.0070 [ 1.0018, 1.0124] | 0.0090 |
| FB group (ref: FB page) | 0.1042 [ 0.0377, 0.2881] | <0.0001 | 0.1542 [ 0.0960, 0.2478] | <0.0001 |
| Intercept | 81.2663 [23.7394, 278.1968] | <0.0001 | 40.6861 [21.3134,77.6674] | <0.0001 |
| *N* | 434,359 | | 1,011,916 | |

*Note*. Results of a negative binomial model estimating the number of reactions of a post as a function of variables shown in the leftmost column. Coef, regression coefficient; IRR, incidence rate ratio; 95% CI, 95% confidence interval of IRR; All standard errors were clustered at the vaccine group level, and all models were estimated with cluster-robust standard error at the vaccine group level.

# S6 Analysis: Exclusive Sources

## S6.1 Exclusive Sources

We identified exclusive sources for pro-vaccine groups and exclusive sources for anti-vaccine groups each month. Exclusive sources for pro-vaccine groups refer to information sources used by five or more pro-vaccine groups and not by any anti-vaccine groups in a given month. For example, if Source A was referenced by seven different pro-vaccine groups but was not referenced or mentioned by any anti-vaccine groups in September 2017, Source A is an exclusive source for pro-vaccine groups of the month. Exclusive sources for anti-vaccine groups, on the other hand, refer to information sources that were used by five or more anti-vaccine groups and not by any pro-vaccine groups in a given month.

We then calculated “source exclusivity.” The source exclusivity of a pro-vaccine [anti-vaccine] group was defined as the proportion of exclusive pro-vaccine [anti-vaccine] sources among all information sources used by the pro-vaccine [anti-vaccine] group in a given month. Source exclusivity ranges from 0% to 100%. For example, if the source exclusivity of an anti-vaccine group was 40%, it means that 4 out of 10 sources used in the group were exclusive sources for anti-vaccine groups, which were not utilized by any pro-vaccine groups in the given month. In Fig. 4 and Fig. S7, the source exclusivity of an anti-vaccine group and the source exclusivity of a pro-vaccine group were referred to as “% exclusive sources in an anti-vaccine group” and “% exclusive sources in a pro-vaccine group,” respectively, for ease of comprehension.

It is worth emphasizing that source exclusivity is the *proportion* of exclusive sources *within* a vaccine group. Source exclusivity is not the total number of exclusive sources summed over vaccine groups. Thus, even if anti-vaccine groups use more information sources in a month or there are more active anti-vaccine groups than active pro-vaccine groups in a month, the average source exclusivity of anti-vaccine groups of the month can be lower than that of pro-vaccine groups.

The average source exclusivities of an anti-vaccine group and a pro-vaccine group each month were displayed in Fig 4A. It is shown that anti-vaccine groups were more dependent on their exclusive sources than pro-vaccine groups. When averaged across the 107 months, the source exclusivity of an anti-vaccine group was 19.3% (*SD* = 4.1%). The source exclusivity of a pro-vaccine group, on the other hand, was only 3.1% (*SD* = 1.5%).

We examined the robustness of these results by calculating source exclusivity for each year, instead of each month. As shown in Fig. S7, when averaged across the 9 years, the source exclusivity of anti-vaccine groups was 15.3% (*SD* = 2.3%), while that of pro-vaccine groups was only 2.9% (*SD* = 1.2%).

We further identified that 2,601 sources were used by at least five anti-vaccine groups but never used by any pro-vaccine groups during the 107-month period. Also, 297 sources were used by at least five pro-vaccine groups but never used by any anti-vaccine groups during the 107-month period. We think that future research will be able to use the lists of these "dedicated" anti-vaccine and pro-vaccine sources in developing automated methods to detect views on vaccines vaccination from online content. The full lists of these sources are available online (*41*).

### S6.1.1 Subtypes Among Exclusive Sources

We identified low credibility, news, social media, and government sources among exclusive sources for anti-vaccine [pro-vaccine] groups each month and defined them as exclusive low credibility sources, exclusive news sources, exclusive social media sources, and exclusive government sources for anti-vaccine [pro-vaccine] groups, respectively.

When averaged over the 107 months (Fig. 4A), exclusive low credibility sources accounted for 5.0% (*SD* = 3.2%) of all exclusive sources used by an anti-vaccine group each month. Exclusive government, news, social media sources were 2.2% (*SD* = 2.5%), 3.3% (*SD* = 3.5%), and 0.1% (*SD* = 0.2%), respectively, among all exclusive sources used by an anti-vaccine group each month.

For pro-vaccine groups (Fig. 4B), exclusive government, news, and social media sources were 3.6% (*SD* = 11.2%), 5.5% (*SD* = 8.0%), and 0.4% (*SD* = 2.5%), respectively, among all exclusive sources used by a pro-vaccine group when averaged over the 107 months. Low credibility sources were not identified among exclusive sources for pro-vaccine groups.

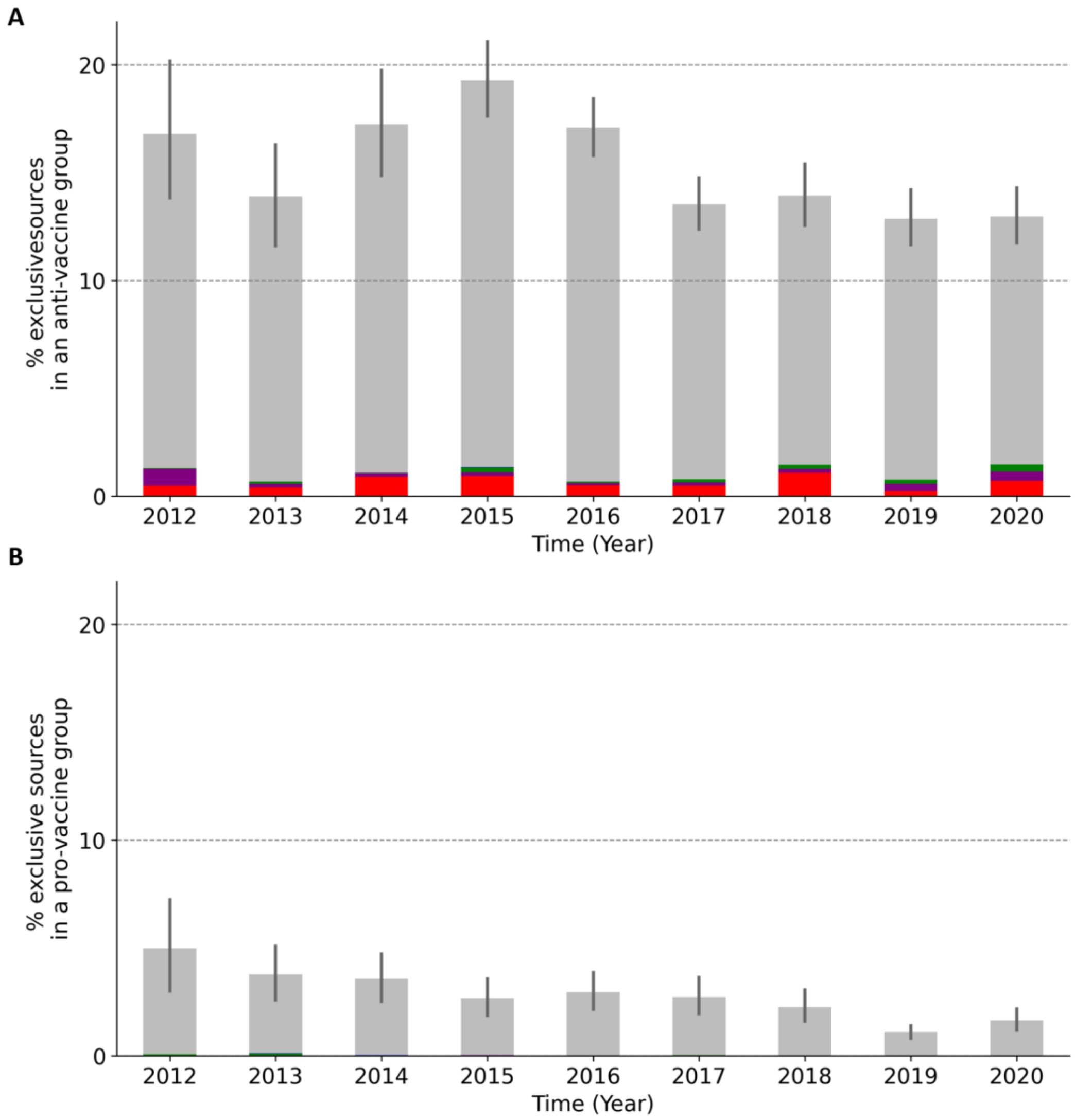

**Fig. S7. Exclusive sources**. **A.** Exclusive sources for anti-vaccine groups by year. The height of each bar indicates the proportion of exclusive sources for anti-vaccine groups among all information sources used by an anti-vaccine group in a year. The red, blue, green, purple, and brown, portions of a bar represent the proportions of exclusive low credibility, social media, government, news, and other sources for anti-vaccine groups, respectively, among all information sources used by an anti-vaccine group. Error bars indicate 95% confidence intervals of the mean proportion of exclusive sources. **B.** Exclusive sources for pro-vaccine groups by year.

# S7 Analysis: Source Popularity and Anti-vaccine Affinity

## S7.1 Source Popularity Among Pro- and Anti-vaccine Groups

We assigned two different popularity scores to each of the 7,605 external information sources, which were used by at least five different pro- or anti-vaccine groups during the 107-month period. These sources originated 91.6% of all URLs linked to external sources of pro- and anti-vaccine groups.

First, the popularity of an information source $i$ among anti-vaccine groups ($r_{ia}$) is calculated as $r_{ia} = n_{ia}/n_a$, where $n_a$ is the total number of posts in anti-vaccine groups using one or more external sources, and $n_{ia}$ is the total number of posts in anti-vaccine groups using $i$ ($0 \leq r_{ia} < 1$). The average $r_{ia}$ was $9.2\times10^{-5}$ and ($SD$ = $1.0\times10^{-3}$).

Second, the popularity of an information source $i$ among pro-vaccine groups ($r_{ip}$) is calculated as $r_{ip} = n_{ip}/n_p$, where $n_p$ is the total number of posts in pro-vaccine groups using one or more external sources, and $n_{ip}$ is the total number of posts in pro-vaccine groups using $i$ ($0 \leq r_{ip} < 1$). The average $r_{ip}$ was $9.0\times10^{-5}$ and ($SD$ = $6.5\times10^{-4}$). Because an external source was used in at least one post, $r_{ia}$ and $r_{ip}$ also meet the following condition: $r_{ia} + r_{ip} > 0$. Due to this condition, the average popularity of a source is always greater than zero: $(r_{ia} + r_{ip})/2 > 0$.

As shown in Fig. S8, there was a negative correlation between $r_{ia}$ and $r_{ip}$ (Spearman $\rho$ = -0.04, $P$ = 0.0004, $N$ = 7605). The correlation remained significant and negative even when the outlier shown in Fig. S8, YouTube, was excluded from the calculation ($\rho$ = -0.04, $P$ = 0.0003, $N$ = 7604). The negative correlations indicate that the more widely used a source is among anti-vaccine groups, the less widely used it is among pro-vaccine groups.

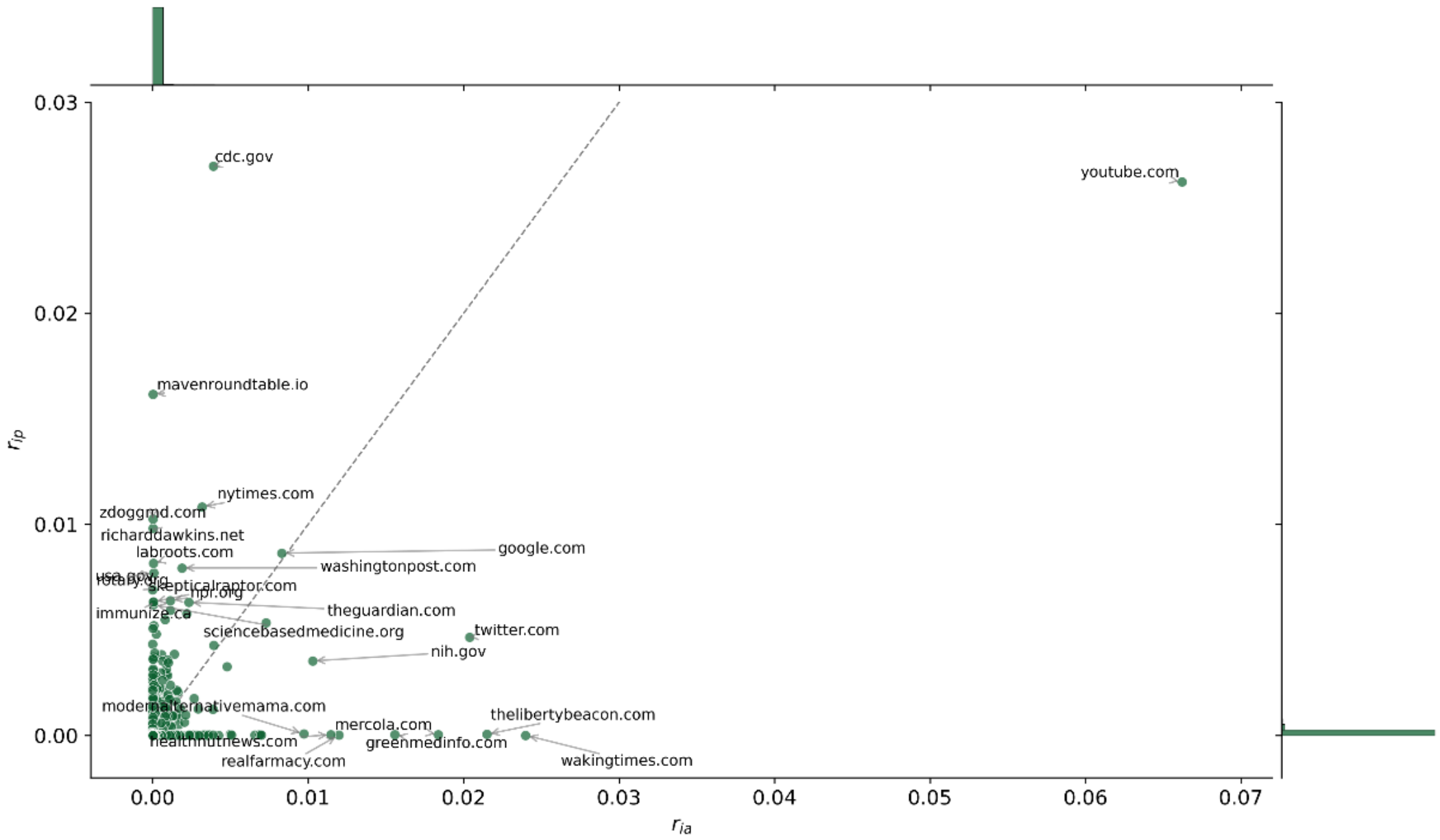

**Fig. S8.** The distribution of $r_{ia}$ and $r_{ip}$ of external sources. The histograms of $r_{ia}$ and $r_{ip}$ on the top and the right, respectively. The dashed line represents the $r_{ia} = r_{ip}$ line. *N* = 7,605.

## S7.2 Anti-vaccine Affinity Score

Using the popularity scores described above, we assigned an anti-vaccine affinity score to each of the external sources used by at least five vaccine groups. The anti-vaccine affinity score of a source represents how widely the source is used by anti-vaccine groups compared with pro-vaccine groups. The anti-vaccine score of Source $i$ is noted $h_{ia}$ and calculated as follows:

$$h_{ia} = \frac{r_{ia} - r_{ip}}{r_{ia} + r_{ip}}$$

This definition of anti-vaccine affinity has several good properties that simplify our analysis and interpretation. First, the score has the maximum value of 1 when Source $i$ is used only by anti-vaccine groups (i.e., $r_{ip} = 0$) and its minimum value of $-1$ when it is used only by pro-vaccine groups (i.e., $r_{ia} = 0$). Second, the score is positive when a source is more popular among anti-vaccine groups (i.e., $r_{ia} > r_{ip}$) and negative when it is more popular among pro-vaccine groups (i.e., $r_{ia} < r_{ip}$). Third, for a source with $r_{ip} > 0$, an anti-vaccine affinity score increases monotonically with a source's relative popularity among anti-vaccine groups, which refers to the ratio between a source's anti-vaccine popularity and its pro-vaccine popularity (i.e., $r_{ia}/r_{ip}$). Thus, for two different sources $i$ and $j$, $h_{ia} < h_{ja}$ if $\frac{r_{ia}}{r_{ip}} < \frac{r_{ja}}{r_{jp}}$.

The distribution of $h_{ia}$ is shown in Fig. S9. The average of $h_{ia}$ was 0.21 (*SD* = 0.75, *N* = 7605).

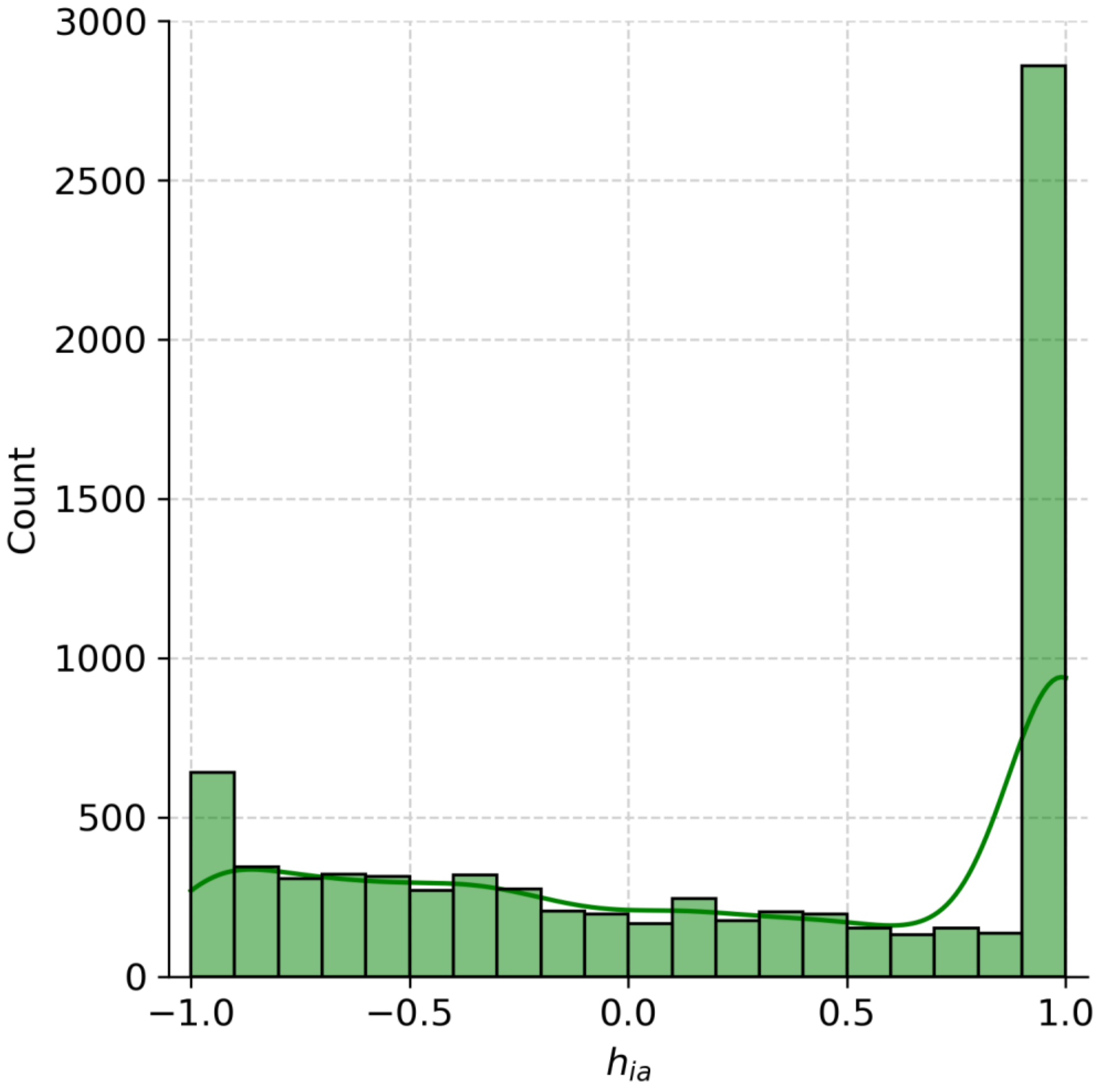

**Fig. S9**. The distribution of $h_{ia}$. The solid line indicates the kernel density estimation. *N* = 7,605.

## S7.3 Ideology Alignment and Anti-vaccine Affinity of Sources

We examined the association between the anti-vaccine affinity score of an information source and the political view that it tends to represent. For this purpose, we adopted a metric reported by Bakshy et al. (*24*) to quantify a source's political ideology. We also checked the robustness of findings using different measures representing a source's political characteristics.

The list reported by Bakshy et al. (*24*) contains the top 500 most shared domains that had been shared at least 44,000 times on Facebook during the period from July 2014 to January 2015. Each domain was assigned an average alignment score, which was calculated by computing the mean political affiliation of Facebook users who shared each news story from each domain and then calculating the average of these mean values for each domain. The authors found that the alignment score well captured the ideological differences among the information sources, so the current study called it the ideology alignment score.

We identified that 329 of the 500 domains reported by Bakshy et al. were used by vaccine groups in the present study. The results are visualized in Fig. S10. In this figure, each circle represents an information source. The color of a circle depends on Source $i$'s anti-vaccine affinity ($h_{ia}$), and the size of a circle is proportional to its average popularity, $(r_{ia} + r_{ip})/2$.

For the 329 domains, we computed the correlation between their ideology alignment scores and anti-vaccine affinity scores. As Fig. S10 demonstrates, there was a highly significant and positive correlation between anti-vaccine affinity and ideology alignment (Spearman $\rho = 0.42$, $P < 0.0001$, $N = 329$), indicating that information sources used more widely by anti-vaccine groups compared with pro-vaccine groups were more ideologically conservative.

The 329 sources displayed in Fig. S10 include two social media sites: twitter.com and youtube.com. Considering the difficulty of defining and determining the overall political characteristics of social media sites (*42*, *43*), we re-estimated the correlation after removing the two social media sites. The result is presented in Fig. 4C, and the correlation was still positive and significant ($\rho = 0.42$, $P < 0.0001$, $N = 327$).

### S7.3.1 Robustness Analysis: Ideology Alignment based on MBFC Data

Media Bias and Fact Check (MBFC, www.mediabiasfactcheck.com) evaluates news media sites and classifies each source into one of the five

categories based on its assessment: left, left-center, center, right-center, and right. In April 2021, we retrieved their evaluation results of 1,920 sources, and 961 sources among them were used by at least five different vaccine groups in our data. An ideology alignment score was assigned to each of the 961 sources based on MBFC's evaluation: -2(left), -1(left-center), 0(center), 1(right-center), 2(right).

We examined if a source's anti-vaccine affinity and its ideology alignment score reported by MBFC are correlated. It was found that there was a significant correlation between anti-vaccine affinity and ideology alignment (Spearman $\rho$ = 0.14, $P < 0.0001$, $N = 961$, Fig. S11), confirming the aforementioned result based on alignment scores from Bakshy et al. (*24*).

### S7.3.2 Robustness Check: Ideology Alignment based on All Sides Data

All Sides (www.allsides.com) publishes the "media bias ratings" that classify media sources into five categories: left, left-center, center, right-center, and right. We obtained All Sides's rating data of 296 news media outlets (*44*). There were 191 sources that existed both in the rating data and our list of external sources used by at least five different vaccine groups. We assigned an ideology alignment score to each of the 191 sources based on All Sides's evaluation: -2(left), -1(left-center), 0(center), 1(right-center), 2(right).

We examined if a source's anti-vaccine affinity and its ideology alignment score based on All Sides' assessment are correlated. It was found that there was a significant and positive correlation between anti-vaccine affinity and ideology alignment (Spearman $\rho$ = 0.47, $P < 0.0001$, $N = 191$, Fig. S12), confirming the findings reported in S7.3 and S7.3.1.

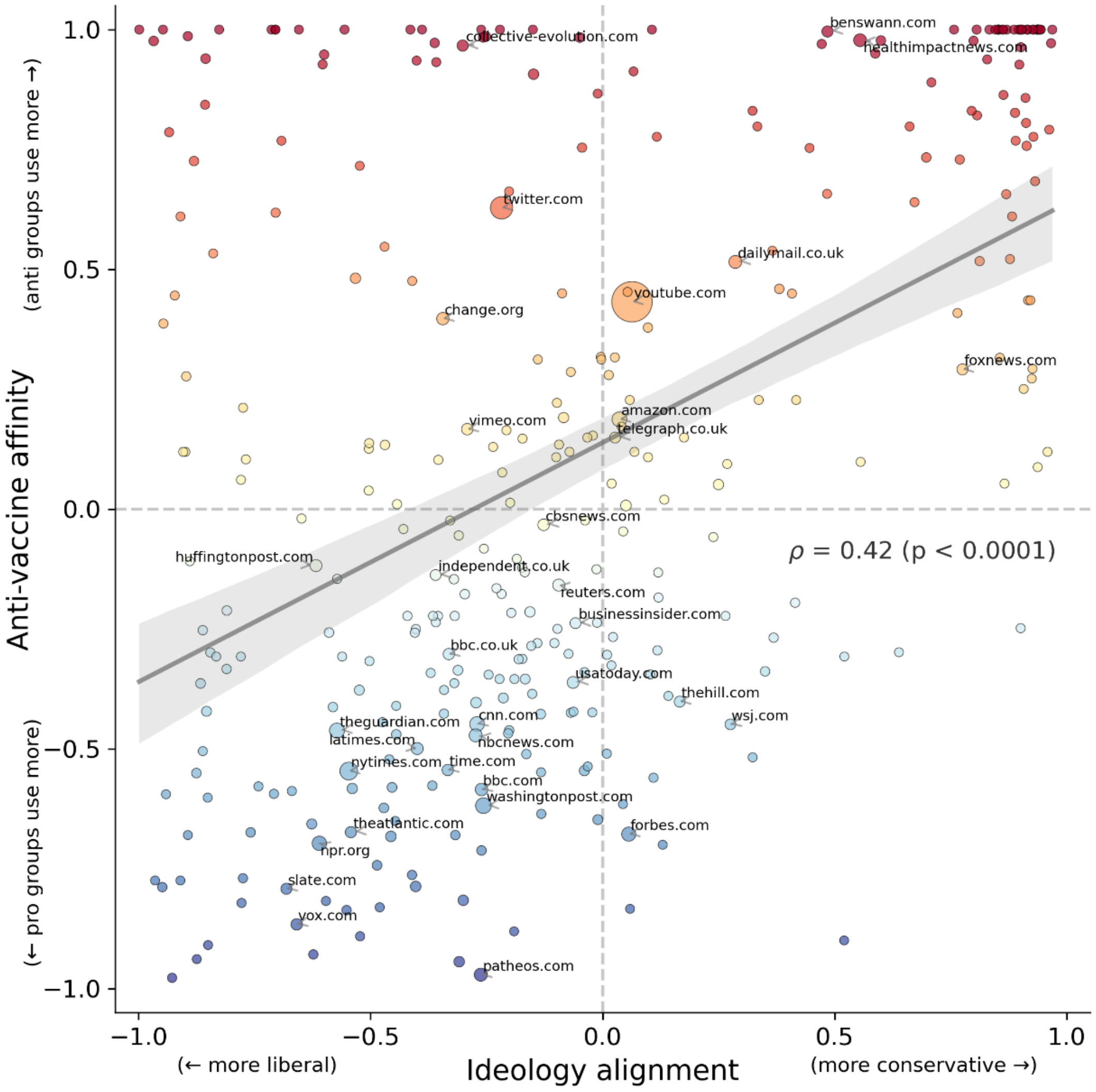

**Fig. S10**. The association between the anti-vaccine affinity of an information source and its ideology alignment reported by Bakshy et al. (24). $N$ = 329.

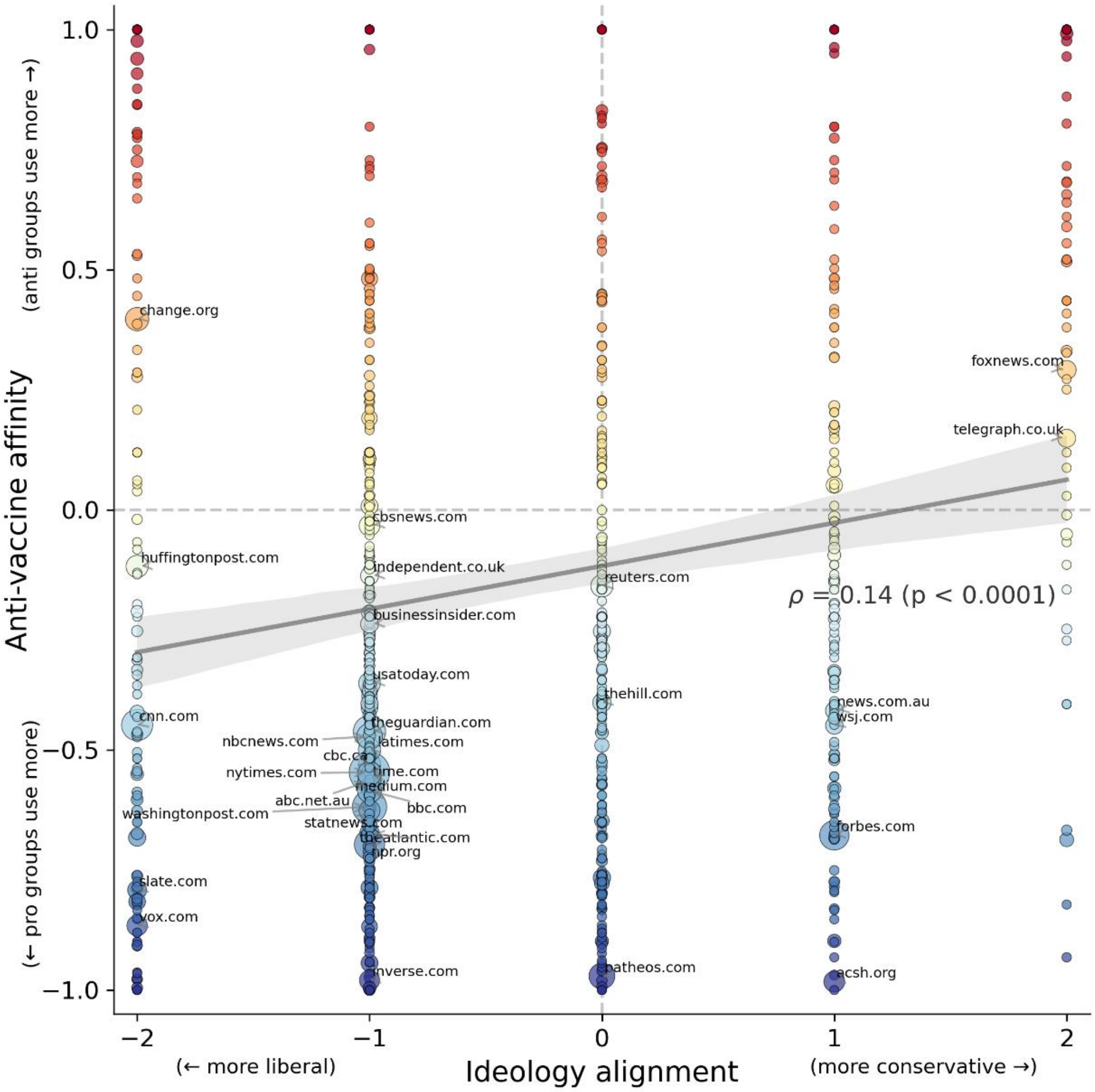

**Fig. S11**. The association between the anti-vaccine affinity of an information source and its ideology alignment reported by MBFC (mediabiasfactcheck.com). $N = 961$.

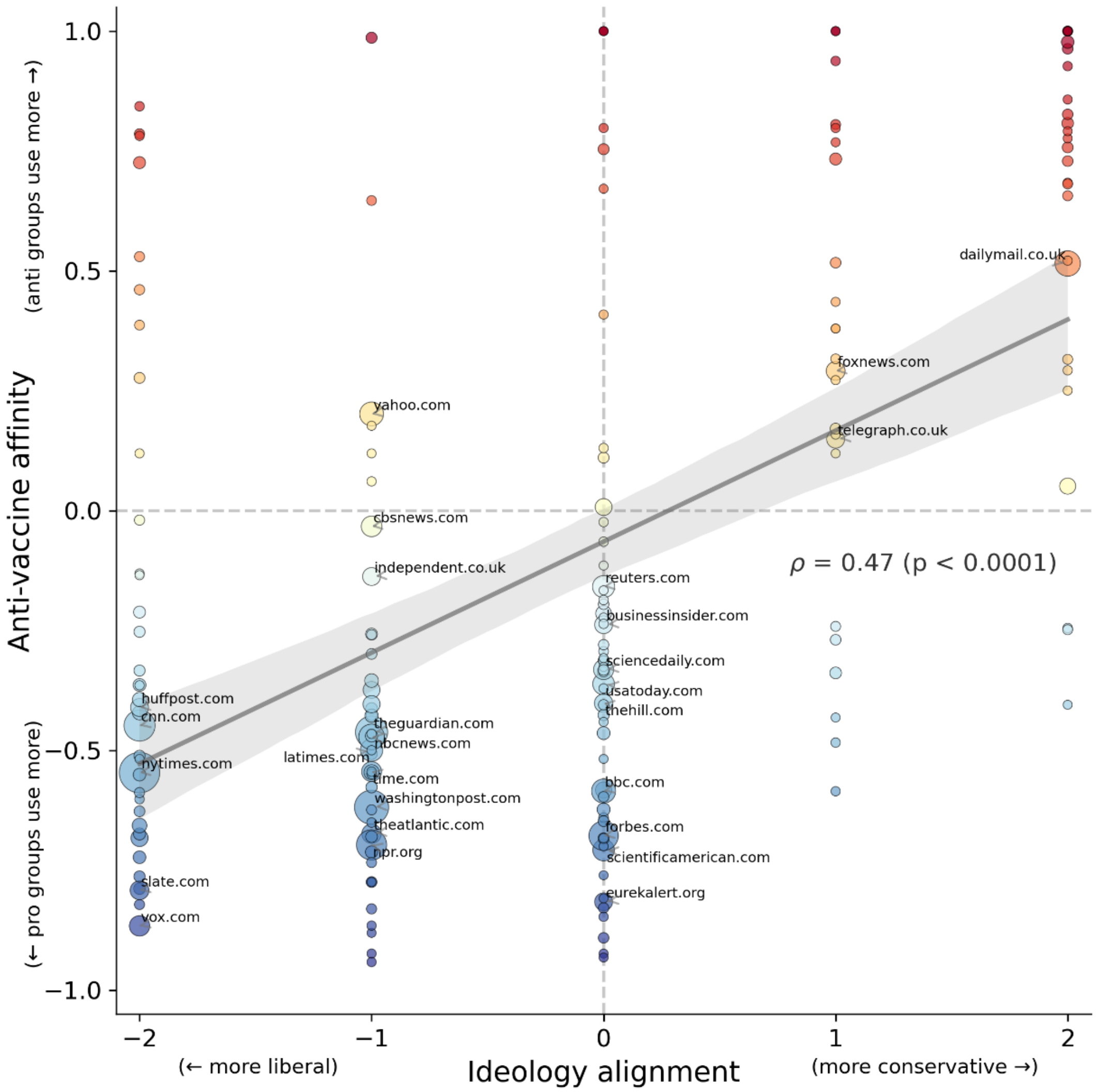

**Fig. S12**. The association between anti-vaccine affinity of an information source and its ideology alignment reported by All Sides (allsides.com). $N = 191$.

## S8 Analysis: Networks of Vaccine Groups within Facebook

We constructed a weighted and undirected network among vaccine groups. In this network, each node represents a vaccine group. An edge connecting two nodes indicates that one vaccine group referenced the other group or its content, and the weight of an edge corresponds to the total number of references between the two groups. All self-loops, which represent a vaccine group's referencing its own content, were removed from the network. Network analysis was conducted using NetworkX (*45*), a Python package for network data analysis. Network visualization was based on Gephi, an open source software for network visualization (*46*). The position of each node in a network diagram was determined by the ForceAtlas2 algorithm (*47*).

Fig. S13 shows the network among 2,328 vaccine groups in our dataset. The size of a node is proportional to its degree. Isolated nodes that are not connected with any other nodes are displayed on the left. The red, blue, and gray circles are anti-vaccine, pro-vaccine, and mixed/neutral vaccine groups, respectively. Edges connecting two anti-vaccine groups; two pro-vaccine groups; an anti-vaccine group and a pro-vaccine group; an anti- or pro-vaccine group with a neutral/mixed vaccine group; and two mixed/neutral groups were colored red, blue, yellow, gray, and gray respectively. In this network, 6.7% of all edges were connecting an anti-vaccine group with a pro-vaccine group, while 66.8% and 26.2% were connecting two anti-vaccine groups and two pro-vaccine groups, respectively.

We analyzed the structure of the network's largest connected component (LCC). The LCC, shown in Fig. 4D, included 1,159 nodes consisting of 410 pro-vaccine, 744 anti-vaccine, and 5 mixed/neutral groups. To examine the level of separation between pro- and anti-vaccine groups, we calculated the proportion of outgroup edges connected to each node (Fig. S14). Outgroup edges refer to edges connecting vaccine groups with different views, such as an edge connecting a pro-vaccine group and an anti-vaccine group. Ingroup edges, on the other hand, refer to edges connecting groups with the same view, such as an edge connecting two pro-vaccine groups. In Fig. S14, red and blue represent anti- and pro-vaccine groups. The height of a bar indicates relative frequency, which ranges from 0 to 1. Solid lines show the result of the Kernel density estimation (left y-axis). Dashed lines indicate the average proportions. As shown in Fig. S14, the proportion of outgroup edges was marginal. Specifically, on average only 11.0% (*SD* = 22.7%, median = 0.0%, *N* = 410) of all edges of a pro-vaccine node and 5.3% (*SD* = 15.6%, median = 0.0%, *N* = 744) of all edges connected to an anti-vaccine node were outgroup edges. The proportion of outgroup edges that an anti-vaccine group had was significantly lower than that of a pro-vaccine group (Mann-Whitney $U_1$ = 128292.5, $P < 0.0001$, $N_1 = 410$, $N_2 = 744$; Kolmogorov-Smirnov $D = 0.192$, $P < 0.0001$). The results imply that pro- and anti-vaccine groups were segregated from each other, and the level of insularity was higher for anti-vaccine groups.

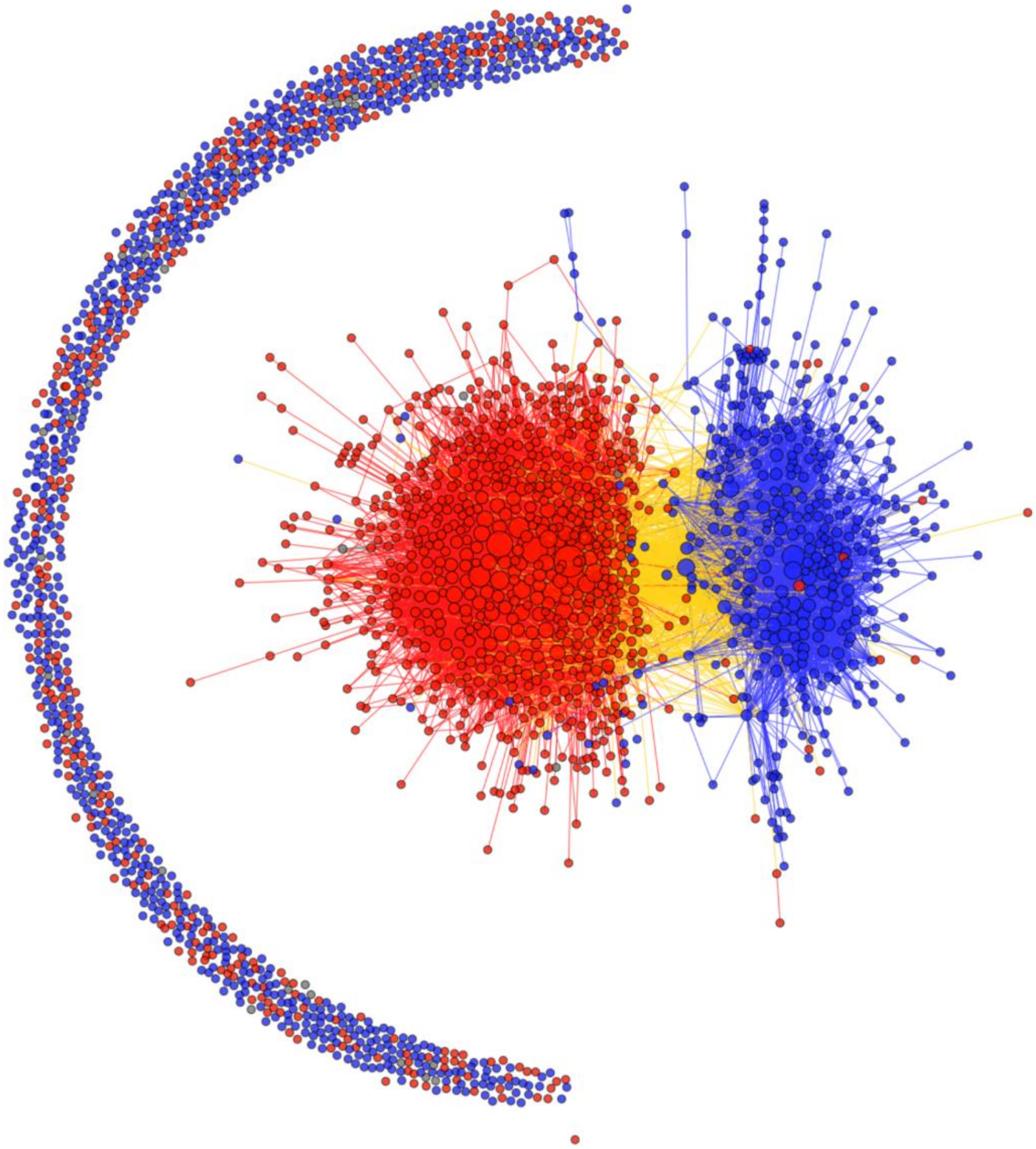

**Fig. S13**. The network of vaccine groups within Facebook.

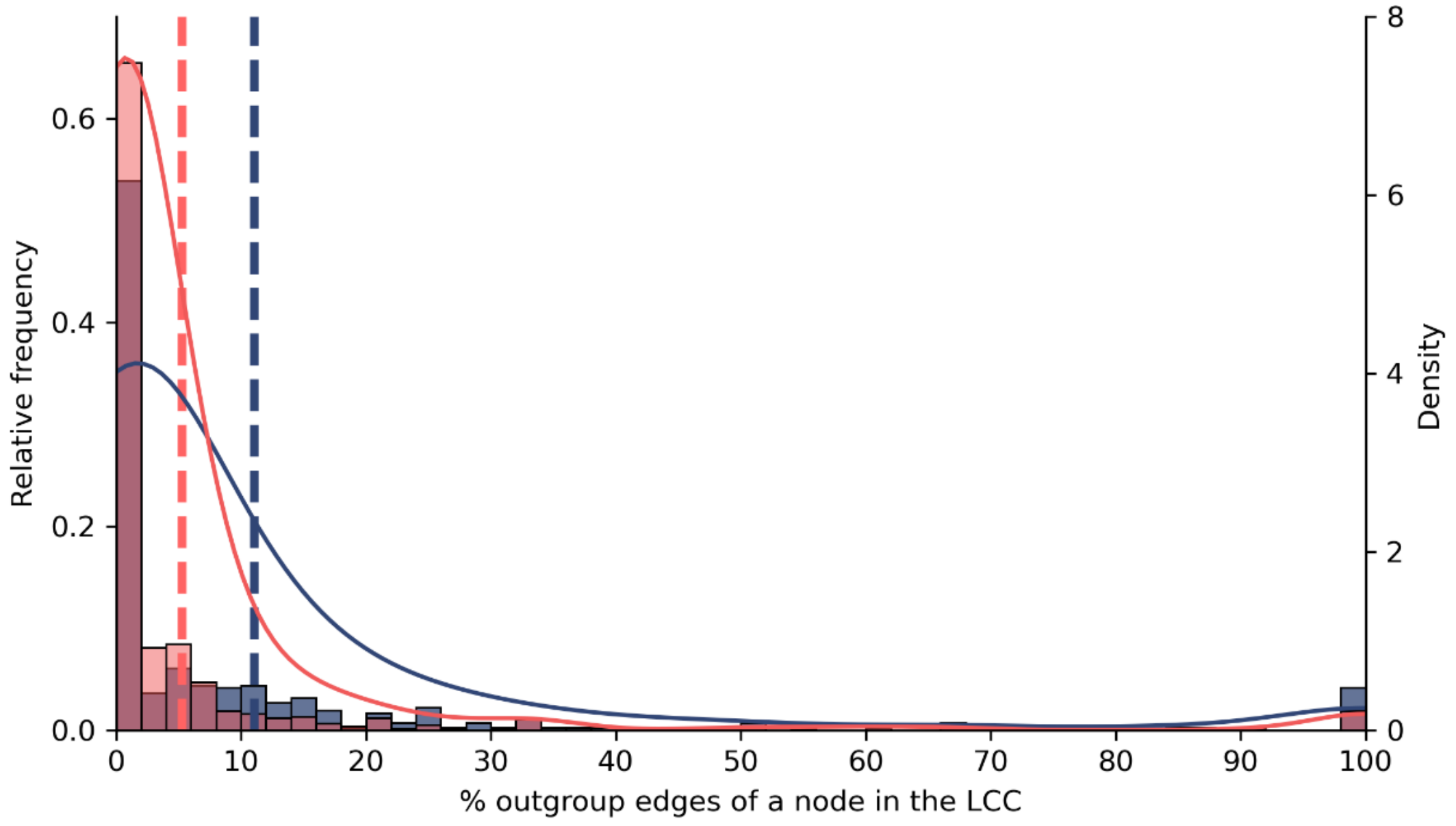

**Fig. S14**. The proportion of outgroup edges of a node in the largest connected component.

## S9 Data and Codes for Replication

Data and codes to reproduce the results in the manuscript are available online (*41*)